\documentstyle[12pt,epsf]{article}
\newlength{\pubnumber} 

\catcode`\@=11
\@addtoreset{equation}{section}

\def\section{\@startsection{section}{1}{\z@}{3.5ex plus 1ex minus .2ex}
 {2.3ex plus .2ex}{\large\bf}}
\def\subsection{\@startsection{subsection}{2}{\z@}{2.3ex plus .2ex}
 {2.3ex plus .2ex}{\bf}}

\def\mbf{\mathbf}
\def\PS{PS  }

\def\LRS{LRS  }
\def\SLM{SLM  }

\def\PSc{PS}
\def\FSUc{FSU5}

\def\SLMc{SLM}

\begin{document}

\begin{titlepage}
\samepage{
\setcounter{page}{1}
\rightline{ACT--09/00, CTP--TAMU--15/00}
\rightline{TPI--MINN--00/34,UMN--TH--1912--00 }
\rightline{OUTP-00--28P}

\rightline{\tt hep-ph/0006331}
\rightline{May 2000}
\vfill
\begin{center}
 {\Large \bf Left--Right Symmetric \\Heterotic--String Derived Models\\}
\vfill
\vfill {\large
	Gerald B. Cleaver$^{1,2}$\footnote{gcleaver@rainbow.physics.tamu.edu},
        Alon E. Faraggi$^{3,4}$\footnote{faraggi@mnhepw.hep.umn.edu}
	and 
	Christopher Savage$^{3}$\footnote{csavage@physics.spa.umn.edu}}\\
\vspace{.12in}
{\it $^{1}$ Center for Theoretical Physics,
	    Texas A\&M University, College Station, TX 77843\\}
\vspace{.05in}
{\it $^{2}$ Astro Particle Physics Group,
            Houston Advanced Research Center (HARC),\\
            The Mitchell Campus,
            Woodlands, TX 77381\\}
\vspace{.05in}
{\it$^{3}$  Department of Physics, University of Minnesota,
            Minneapolis, MN 55455\\}
\vspace{.05in}
{\it$^{4}$  Theoretical Physics Department, University of Oxford,
            Oxford, OX1 3NP\\}
\vspace{.025in}
\end{center}
\vfill
\begin{abstract}
Recently it was demonstrated that free fermionic heterotic--strings
can produce models with solely the Minimal Supersymmetric
Standard Model states in the low energy spectrum. This unprecedented
result provides further strong evidence for the possibility that the
true string vacuum shares some of the properties of the free fermionic
models. Past free fermionic models have focused on several possible
unbroken observable $SO(10)$ subgroups at the string scale, which include
the flipped $SU(5)$ (FSU5), the Pati--Salam (PS) string models, 
and the string Standard--like Models (SLM). We extend this study
to include the case in which the $SO(10)$ symmetry is broken to
the Left--Right Symmetric (LRS) gauge group, $SO(10)\rightarrow
SU(3)_C\times U(1)_{B-L}\times SU(2)_L\times SU(2)_R$. 
We present several models of this type and discuss their
phenomenological features. The most striking new outcome of the \LRS
string models, in contrast to the case of the \FSUc, the \PSc, and the
\SLM string models, is that they can produce effective field theories 
that are free of Abelian anomalies. We discuss the distinction between
the two types of free fermionic models which result in the presence,
or absence, of an anomalous $U(1)$. As a counter
example we also  present a \LRS model that does contain an anomalous
$U(1)$. Additionally, we discuss how in string models the Standard Model
spectrum may arise from the three $\mbf 16$ representations of $SO(10)$, 
while the weak--hypercharge does not have the canonical $SO(10)$ 
embedding.
\end{abstract}
\smallskip}
\end{titlepage}

\setcounter{footnote}{0}

\def\beq{\begin{equation}}
\def\eeq{\end{equation}}
\def\beqn{\begin{eqnarray}}
\def\eeqn{\end{eqnarray}}

\def\no{\noindent }
\def\nolabel{\nonumber }
\def\ie{{\it i.e.}}
\def\eg{{\it e.g.}}
\def\half{{\textstyle{1\over 2}}}
\def\third{{\textstyle {1\over3}}}
\def\quarter{{\textstyle {1\over4}}}
\def\sixth{{\textstyle {1\over6}}}
\def\m{{\tt -}}
\def\p{{\tt +}}

\def\Tr{{\rm Tr}\, }
\def\tr{{\rm tr}\, }

\def\slash#1{#1\hskip-6pt/\hskip6pt}
\def\slk{\slash{k}}
\def\GeV{\,{\rm GeV}}
\def\TeV{\,{\rm TeV}}
\def\y{\,{\rm y}}
\def\SM{Standard--Model }
\def\SUSY{supersymmetry }
\def\SSSM{supersymmetric standard model}
\def\vev#1{\left\langle #1\right\rangle}
\def\l{\langle}
\def\r{\rangle}
\def\o#1{\frac{1}{#1}}

\def\Htw{{\tilde H}}
\def\chibar{{\overline{\chi}}}
\def\qbar{{\overline{q}}}
\def\ibar{{\overline{\imath}}}
\def\jbar{{\overline{\jmath}}}
\def\Hbar{{\overline{H}}}
\def\Qbar{{\overline{Q}}}
\def\abar{{\overline{a}}}
\def\alphabar{{\overline{\alpha}}}
\def\betabar{{\overline{\beta}}}
\def\tautwo{{ \tau_2 }}
\def\thetatwo{{ \vartheta_2 }}
\def\thetathree{{ \vartheta_3 }}
\def\thetafour{{ \vartheta_4 }}
\def\ttwo{{\vartheta_2}}
\def\tthree{{\vartheta_3}}
\def\tfour{{\vartheta_4}}
\def\ti{{\vartheta_i}}
\def\tj{{\vartheta_j}}
\def\tk{{\vartheta_k}}
\def\calF{{\cal F}}
\def\smallmatrix#1#2#3#4{{ {{#1}~{#2}\choose{#3}~{#4}} }}
\def\ab{{\alpha\beta}}
\def\Minv{{ (M^{-1}_\ab)_{ij} }}
\def\bone{{\bf 1}}
\def\bo{{\bf 1}}
\def\ii{{(i)}}
\def\V{{\bf V}}
\def\N{{\bf N}}

\def\bfb{{\bf b}}
\def\bfS{{\bf S}}
\def\bfX{{\bf X}}
\def\bfI{{\bf I}}
\def\ma{{\mathbf a}}
\def\mb{{\mathbf b}}
\def\mS{{\mathbf S}}
\def\mX{{\mathbf X}}
\def\mI{{\mathbf I}}
\def\malpha{{\mathbf \alpha}}
\def\mbeta{{\mathbf \beta}}
\def\mgamma{{\mathbf \gamma}}
\def\mzeta{{\mathbf \zeta}}
\def\mxi{{\mathbf \xi}}

\def\t#1#2{{ \Theta\left\lbrack \matrix{ {#1}\cr {#2}\cr }\right\rbrack }}
\def\C#1#2{{ C\left\lbrack \matrix{ {#1}\cr {#2}\cr }\right\rbrack }}
\def\tp#1#2{{ \Theta'\left\lbrack \matrix{ {#1}\cr {#2}\cr }\right\rbrack }}
\def\tpp#1#2{{ \Theta''\left\lbrack \matrix{ {#1}\cr {#2}\cr }\right\rbrack }}
\def\l{\langle}
\def\r{\rangle}

\def\cL#1#2{{\cal L}^{#1}_{#2}} 
\def\x#1{\phi_{#1}}
\def\bx#1{{\bar{\phi}}_{#1}}

\def\cl#1{{\cal L}_{#1}} 
\def\bcl#1{\bar{\cal L}_{#1}} 

\def\bt{{\bar 3}}
\def\h#1{h_{#1}}
\def\Q#1{Q_{#1}}
\def\L#1{L_{#1}}

\def\N#1{N_{#1}}
\def\bN#1{{\bar{N}}_{#1}}

\def\S#1{S_{#1}}
\def\Ss#1#2{S_{#1}^{#2}}
\def\bS#1{{\bar{S}}_{#1}}
\def\bSs#1#2{{\bar{S}}_{#1}^{#2}}

\def\D#1{D_{#1}}
\def\Ds#1#2{D_{#1}^{#2}}
\def\bD#1{{\bar{D}}_{#1}}
\def\bDs#1#2{{\bar{D}}_{#1}^{#2}}

\def\p#1{\phi_{#1}}
\def\bp#1{{\bar{\phi}}_{#1}}

\def\P#1{\Phi_{#1}}
\def\bP#1{{\bar{\Phi}}_{#1}}
\def\X#1{\Phi_{#1}}
\def\bX#1{{\bar{\Phi}}_{#1}}
\def\Ps#1#2{\Phi_{#1}^{#2}}
\def\bPs#1#2{{\bar{\Phi}}_{#1}^{#2}}
\def\ps#1#2{\phi_{#1}^{#2}}
\def\bps#1#2{{\bar{\phi}}_{#1}^{#2}}
\def\php{\phantom{+}}

\def\H#1{H_{#1}}
\def\bH#1{{\bar{H}}_{#1}}

\def\xH#1#2{H^{#1}_{#2}}
\def\bxH#1#2{{\bar{H}}^{#1}_{#2}}


\def\inbar{\,\vrule height1.5ex width.4pt depth0pt}

\def\IC{\relax\hbox{$\inbar\kern-.3em{\rm C}$}}
\def\IQ{\relax\hbox{$\inbar\kern-.3em{\rm Q}$}}
\def\IR{\relax{\rm I\kern-.18em R}}
 \font\cmss=cmss10 \font\cmsss=cmss10 at 7pt
\def\IZ{\relax\ifmmode\mathchoice
 {\hbox{\cmss Z\kern-.4em Z}}{\hbox{\cmss Z\kern-.4em Z}}
 {\lower.9pt\hbox{\cmsss Z\kern-.4em Z}}
 {\lower1.2pt\hbox{\cmsss Z\kern-.4em Z}}\else{\cmss Z\kern-.4em Z}\fi}

\def\AEF{A.E. Faraggi}
\def\NPB#1#2#3{{\it Nucl.\ Phys.}\/ {\bf B#1} (#2) #3}
\def\PLB#1#2#3{{\it Phys.\ Lett.}\/ {\bf B#1} (#2) #3}
\def\PRD#1#2#3{{\it Phys.\ Rev.}\/ {\bf D#1} (#2) #3}
\def\PRL#1#2#3{{\it Phys.\ Rev.\ Lett.}\/ {\bf #1} (#2) #3}
\def\PRT#1#2#3{{\it Phys.\ Rep.}\/ {\bf#1} (#2) #3}
\def\MODA#1#2#3{{\it Mod.\ Phys.\ Lett.}\/ {\bf A#1} (#2) #3}
\def\IJMP#1#2#3{{\it Int.\ J.\ Mod.\ Phys.}\/ {\bf A#1} (#2) #3}
\def\nuvc#1#2#3{{\it Nuovo Cimento}\/ {\bf #1A} (#2) #3}
\def\RPP#1#2#3{{\it Rept.\ Prog.\ Phys.}\/ {\bf #1} (#2) #3}
\def\etal{{\it et al\/}}

\hyphenation{su-per-sym-met-ric non-su-per-sym-met-ric}
\hyphenation{space-time-super-sym-met-ric}
\hyphenation{mod-u-lar mod-u-lar--in-var-i-ant}


\setcounter{footnote}{0}
\section{Introduction}
Recently it was demonstrated that free fermionic heterotic 
string models can produce models with solely the spectrum of the
Minimal Supersymmetric Standard Model (MSSM) in the effective
four dimensional field theory \cite{cfn1}. This achievement
provides further motivation to improve our understanding of
this particular class of heterotic string models. 
The realistic free fermionic models consist of a large number of
three generation models which differ in their detailed 
phenomenological characteristics. All these models share an
underlying $\IZ_2\times \IZ_2$ orbifold structure which arises
{}from a basic set of boundary condition basis vectors, 
the so--called {\bf NAHE}--set\footnote{NAHE=pretty in Hebrew. The NAHE
set was first employed by Nanopoulos, Antoniadis, Hagelin and Ellis
in the construction of the flipped $SU(5)$ heterotic--string model
\cite{flipped}. Its vital role in the realistic free fermionic models has been
emphasized in ref.~\cite{nahe}.}.
With this fundamental set incorporated as a necessary ingredient
in the construction, one then finds that three generation models,
with the canonical $SO(10)$ embedding of the Standard Model 
spectrum\footnote{It is interesting to note that among the perturbative
heterotic--string orbifold models the free fermionic models
are the only ones which have yielded three generations with the 
canonical $SO(10)$ embedding} naturally arise. Furthermore,
one of the generic features of semi--realistic string
constructions is the existence of numerous massless states
beyond the MSSM spectrum, some of which carry fractional electric
charge and hence must decouple from the low energy spectrum.
Recently, and for the first time since the advent of string phenomenology
\cite{candelas}, we have been able to demonstrate in the FNY free fermionic
model \cite{fny}, that free fermionic models can also produce models with 
solely the MSSM states in the light spectrum. 
We will refer to such a heterotic string model, as a Minimal
Standard Heterotic String Model (MSHSM)\footnote{
It should be emphasized that the success of the 
FNY model in producing a MSHSM should not be regarded
as indicating that the FNY model represents the correct
string vacuum. Indeed, much further elaborate studies
would be needed to support such a claim.
The phenomenological success of the free fermionic models 
implies that the generic structure afforded by the NAHE set
is favorable for obtaining agreement with the phenomenological
characteristics suggested by the Standard Model data.}.

At the level of the NAHE set, denoted by $\{{\bf 1},\mS,\mb_1,\mb_2,\mb_3\}$, 
the gauge group is $SO(10)\times SO(6)^3\times E_8$.
The $SO(6)^3$ symmetries are horizontal
flavor symmetries; the $E_8$ factor gives rise to the hidden
gauge group at this stage and the Standard Model universal gauge group
arises from the $SO(10)$ factor.
Beyond the NAHE set the construction of the realistic free fermionic
models proceeds by adding three or four additional boundary condition
basis vectors. These additional basis vectors fix the final $SO(10)$
subgroup in the effective field theory, and at the same time
reduce the number of generations to three, one from 
each of the sectors $\mb_1$, $\mb_2$ and $\mb_3$. The models studied
to date have focused on three possibilities for the final $SO(10)$
subgroup: the flipped SU(5) (FSU5) with 
$SO(10)\rightarrow SU(5)\times U(1)$ \cite{flipped};
the Pati--Salam (PS) type models with $SO(10)\rightarrow SO(6)\times SO(4)$
\cite{alr}; and the standard--like models (SLM)
with $SO(10)\rightarrow SU(3)\times SU(2)\times U(1)^2$ \cite{slm}.

In this paper we extend the analysis of the three generation free
fermionic models to models with the left--right symmetric (LRS) gauge
group, {\it i.e.}~with $SO(10)\rightarrow SU(3)\times U(1)
\times SU(2)_L\times SU(2)_R$. Our primary motivation is to extend
our understanding of the general properties of the realistic
free fermionic models. 
It should also be noted, however,
that \LRS  models have been extensively
studied as attractive field theoretic extensions of the 
Standard Model, in which parity violation is understood to arise
{}from the spontaneous breakdown of $SU(2)_R$. Further phenomenological
advantages of \LRS models include its potential
role in providing a solution to the strong CP problem and 
to the SUSY CP problem \cite{rabi}. From a supersymmetric grand unification
perspective the \LRS symmetric models have the appealing
property that R--parity appears as a gauged symmetry. From
a string unification perspective the \LRS
models, similar to the \PS and \SLM string models,
have the advantage that they can incorporate the stringy doublet--triplet
splitting mechanism \cite{ps}. In contrast to the MSSM, and similar to 
the \PS models, the \LRS models 
produce Yukawa couplings of the up and down quark families
to a Higgs bi--doublet, which present the danger of
inducing Flavor Changing Neutral Currents (FCNC) at
an unacceptable rate \cite{lrsfcnc}. 

One of the interesting aspects of the \LRS string models
that we show is the possible absence of an anomalous $U(1)$ symmetry. 
As is well known, generically string models with (2,0) world--sheet
supersymmetry give rise to an anomalous $U(1)$ symmetry \cite{bs,kn,cf1}. 
In this paper we present the first examples of semi--realistic (2,0)
heterotic string models in which all the $U(1)$ symmetries
are anomaly free. 

Our paper is organized as follows. In Section \ref{leftrightft}
we give a quick review of the field theory content of the \LRS 
models that we aim to construct. In Section \ref{leftrightffm}
we discuss the symmetry breaking pattern in the string models. In Section
\ref{noes} we present the first example of a \LRS
string model. The full massless spectrum and related quantum numbers
with respect to the four dimensional gauge group are determined, 
as well as all superpotential terms up to quintic order.
In Section \ref{es} we offer a variation of the \LRS
model in which the $U(1)_{B-L}$ symmetry
is enhanced to a non--Abelian symmetry. 
The full massless spectrum with quantum numbers and the superpotential 
are derived for this model as well. In Section \ref{anomalousu1}
we discuss the absence of an anomalous $U(1)$ symmetry in
our first two examples of \LRS  string models.
As a counter example we also present a \LRS model
which does possess an anomalous $U(1)$. Section \ref{conclusion}
contains a phenomenological discussion and our conclusions. 
 
\setcounter{footnote}{0}
\section{The supersymmetric left--right symmetric model}\label{leftrightft}

In this section we briefly summarize the field theory
structure of the type of models that we aim to construct from 
string theory in this paper. The observable sector gauge symmetry 
we seek is 
$SU(3)_C\times SU(2)_L\times SU(2)_R\times U(1)_{B-L}$. 
Such models are reminiscent of the \PS
type string models, but differ from them by the fact that the $SU(4)$
gauge group is broken to $SU(3)\times U(1)_{B-L}$ already at the 
string level. Similar to the \PS models, the \LRS
models possess the $SO(10)$ embedding. 
The quarks and leptons are accommodated in the following
representations:
\beqn
Q_L^{i} &=& (3,2,1)_{1\over6}  ~~=~ {u\choose d}^i\\
Q_R^{i} &=& ({\bar 3},1,2)_{-{1\over6}} ~=~  {d^{c}\choose u^{c}}^i\\
L_L^{i} &=& (1,2,1)_{-{1\over2}} ~=~  {\nu\choose e}^i\\
L_R^{i} &=& (1,1,2)_{1\over2} ~~=~  {e^c\choose \nu^c}^i\\
h       &=& (1,2,2)_0 ~~=~  
{\left(\matrix{
                 h^u_+  &  h^d_0\cr
                 h^u_0  &  h^d_-\cr}\right)}
\label{LRsymreps}
\eeqn
where $h^d$ and $h^u$ are the two low energy supersymmetric superfields
associated with the Minimal Supersymmetric Standard Model. The breaking
of $SU(2)_R$ could be achieved with the VEV of $h$. However, this
will result with too light $W_R^\pm$ gauge boson masses. Additional
fields that can be used to break $SU(2)_R$ must therefore be postulated.
The simplest set would consist of two fields $H+\overline{H}$
transforming as $(1,1,2)_{-{1\over2}}+(1,1,{\bar 2})_{1\over2}$.
When $H$ and $\overline{H}$ acquire VEVs along their neutral
components $SU(2)_R\times U(1)_{B-L}$ is broken to the 
Standard Model weak--hypercharge, $U(1)_Y$.
With this symmetry breaking pattern the bi--doublet
Higgs field may split into the two Higgs doublet $h^u$ and
$h^d$ of the MSSM. 

The \LRS string models can also contain Higgs fields that
transform as $(3,1,1)$ and $({\bar 3},1,1)$, which originate from
the vectorial {\bf 10} representation of $SO(10)$. These color triplets
mediate proton decay through dimension five operators, and 
consequently must be sufficiently heavy to insure agreement with
the proton lifetime. An important
advantage of the \LRS breaking pattern,
with $SO(10)\rightarrow SO(6)\times SO(4)$ at the string construction level, 
is that these color triplets may be projected out by the GSO
projections, and therefore need not be present in the
low energy spectrum. In the \PS models, however, 
the Higgs representations that induce $SU(4)\times SU(2)_R\rightarrow
SU(3)_C\times U(1)_{Y}$ contain Higgs triplet representations. 
In the supersymmetric \PS models the color triplets
in the vectorial representation $(6,1,1)$ are used to give
large mass to the Higgs color triplets, by the superpotential
terms $\lambda_2HHD+\lambda_3 {\bar H}{\bar H}{\bar D}$, when the fields
$H$ and ${\bar H}$ develop a large VEV of the order of the GUT scale.
Therefore, the stringy doublet--triplet splitting mechanism is useful
only in models with $SU(3)_C\times SU(2)_L\times U(1)^2$, or 
$SU(3)_C\times SU(2)_L\times SU(2)_R \times U(1)_{B-L}$, as the $SO(10)$ subgroup
which remains unbroken by the GSO projections. 

The \LRS models should also contain four additional
singlet fields $\phi_0$ and $\phi_{i=1,2,3}$. $\phi_0$ acquires
a VEV of the order of the electroweak scale which induces the 
electroweak Higgs doublet mixing, while $\phi_i$ are used 
to construct an extended see--saw mechanism which generate
left--right Majorana masses for the left--handed neutrinos.
The tree level superpotential of the model is given by:
\beq
W=\lambda^1_{ij}Q_L^iQ_R^jh+\lambda^2_{ij}L_L^iL_R^jh+
\lambda^3_{ij}L_R^i{\bar H}\phi^j+
\lambda^4hh\phi^0+\lambda_5\Phi^3
\label{superpot}
\eeq
where $\Phi=\{\phi^i,\phi^0\}$. The superpotential in
eq.~(\ref{superpot}) leads to the neutrino mass matrix
\beq
{\left(\matrix{
                 0 & m_u^{ij}        &  0           \cr
         m_u^{ji}  &   0             &  \l{\bar H}\r\cr
                 0 &   \l{\bar H}\r  &  \l\phi_0\r}\right)}
\label{seesaw}
\eeq
whose diagonalization gives three light neutrinos with masses
of the order $\l\phi_0\r(m_u^{ij}/\l{\bar H}\r)^2$ and gives 
heavy mass, of order $\l{\bar H}\r$, to the right--handed neutrinos.

Below the scale of $SU(2)_R$ breaking the left--right symmetric models
should reproduce the spectrum and couplings of the MSSM. As our interest
here is primarily in the string construction of left--right symmetric 
models we do not enter into the field theory details, which have 
been amply studied in the literature \cite{rabi}. 
There is one important issue, however, that deserves mention. 
As seen from eq.~(\ref{superpot}) in the left--right symmetric models
both up--quark and down--quark masses arise from the coupling to the
Higgs bi--doublet. This introduces the danger of inducing Flavor 
Changing Neutral Currents (FCNC) at an unacceptable rate. A possible 
solution is to use two bi--doublet Higgs representations, 
one of which is used to give masses to the up--type quarks, while
the second is used to give masses to the down--type quarks. 
This, however, introduces a bi--doublet splitting problem. 
Namely, we must insure that one Higgs multiplet 
remains light to give mass to the up-- or down--type quarks, while
the second Higgs multiplet in the respective bi--doublet
becomes sufficiently heavy so as to avoid problems with FCNC.
Arguably, this can be achieved in a field theory setting. However, 
the bi--doublet splitting mechanisms that have been discussed
in the literature \cite{bidoublet}
utilize $SU(2)$ triplet representations that are,
in general, not present in the free fermionic string models.
Therefore, whether
or not bi--doublet splitting can be achieved in the left--right symmetric
string models is an open question, 
which we will not address in this paper. 

We emphasize that our intent here is not to construct a fully 
realistic left--right symmetric model, but merely to study the structure of 
free fermionic string models with this choice of the $SO(10)$ 
subgroup. In this respect we note that the bi--doublet
splitting problem introduces further motivation for
the choice of $SU(3)\times SU(2)\times U(1)^2$ as
the $SO(10)$ subgroup which remains unbroken after 
application of the string GSO projections. 
Thus, while the doublet--triplet splitting problem 
does not distinguish between the \PS string model ($SO(10)\rightarrow
SO(6)\times SO(4)$), or \LRS string model
($SO(10)\rightarrow SU(3)\times SU(2)^2\times U(1)$),
and the \SLM string model ($SO(10)\rightarrow SU(3)\times
SU(2)\times U(1)^2$), the bi--doublet splitting problem
favors the later choice. The \SLM string models
provide a stringy solution both to the doublet--triplet splitting
problem, as well as the bi--doublet splitting problem. 
In this respect it should also be noted that the choice
of $SU(4)_C\times SU(2)_L\times U(1)_{T_{3_R}}$ as the unbroken
$SO(10)$ subgroup also achieves these two tasks. Study of this
case is left for future work. 

\setcounter{footnote}{0}
\section{Left--right symmetric free fermionic models}\label{leftrightffm}

A model in the free fermionic formulation \cite{fff} is constructed by 
choosing a consistent set of boundary condition basis vectors.
The basis vectors, $\mb_k$, span a finite  
additive group $\Xi=\sum_k{{n_k}{\mb_k}}$
where $n_k=0,\cdots,{{N_{z_k}}-1}$.
The physical massless states in the Hilbert space of a given sector
$\malpha\in{\Xi}$, are obtained by acting on the vacuum with 
bosonic and fermionic operators and by
applying the generalized GSO projections. The $U(1)$
charges, $Q(f)$, for the unbroken Cartan generators of the four 
dimensional gauge group are in one 
to one correspondence with the $U(1)$
currents ${f^*}f$ for each complex fermion f, and are given by:
\begin{equation}
{Q(f) = {1\over 2}\alpha(f) + F(f)},
\label{u1charges}
\end{equation}
where $\alpha(f)$ is the boundary condition of the world--sheet fermion $f$
in the sector $\malpha$, and 
$F_\alpha(f)$ is a fermion number operator counting each mode of 
$f$ once (and if $f$ is complex, $f^*$ minus once). 
For periodic fermions,
$\alpha(f)=1$, the vacuum is a spinor representation of the Clifford
algebra of the corresponding zero modes. 
For each periodic complex fermion $f$
there are two degenerate vacua ${\vert +\rangle},{\vert -\rangle}$ , 
annihilated by the zero modes $f_0$ and
${{f_0}^*}$ and with fermion numbers  $F(f)=0,-1$, respectively. 

The realistic models in the free fermionic formulation are generated by 
a basis of boundary condition vectors for all world--sheet fermions 
\cite{flipped,fny,alr,slm,eu,custodial}. The basis is constructed in
two stages. The first stage consists of the NAHE set \cite{nahe,slm}, 
which is a set of five boundary condition basis 
vectors, $\{{{\bf 1},\mS,\mb_1,\mb_2,\mb_3}\}$. 
The gauge group after the NAHE set
is $SO(10)\times SO(6)^3\times E_8$ with $N=1$ space--time supersymmetry. 
The vector $\mS$ is the supersymmetry generator and the superpartners of
the states from a given sector $\malpha$ are obtained from the sector
$\mS+\malpha$. The space--time vector bosons that generate the gauge group 
arise from the Neveu--Schwarz (NS) sector and from the sector $\mzeta
\equiv {\mathbf 1}+\mb_1+\mb_2+\mb_3$.
The NS sector produces the generators of 
$SO(10)\times SO(6)^3\times SO(16)$. The sector 
$\mzeta$
produces the spinorial $\bf 128$ of $SO(16)$ and completes the hidden 
gauge group to $E_8$. The vectors $\mb_1$, $\mb_2$ and $\mb_3$ 
produce 48 spinorial $\bf 16$'s of $SO(10)$, sixteen from each sector $\mb_1$, 
$\mb_2$ and $\mb_3$. The vacuum of these sectors contains eight periodic 
worldsheet fermions, five of which produce the charges under the 
$SO(10)$ group, while the remaining three periodic fermions 
generate charges with respect to the flavor symmetries. Each of the 
sectors $\mb_1$, $\mb_2$ and $\mb_3$ is charged with respect to a 
different set of flavor quantum numbers, $SO(6)_{1,2,3}$.

The NAHE set divides the 44 right--moving and 20 left--moving real internal
fermions in the following way: ${\bar\psi}^{1,\cdots,5}$ are complex and
produce the observable $SO(10)$ symmetry; ${\bar\phi}^{1,\cdots,8}$ are
complex and produce the hidden $E_8$ gauge group;
$\{{\bar\eta}^1,{\bar y}^{3,\cdots,6}\}$, $\{{\bar\eta}^2,{\bar y}^{1,2}
,{\bar\omega}^{5,6}\}$, $\{{\bar\eta}^3,{\bar\omega}^{1,\cdots,4}\}$
give rise to the three horizontal $SO(6)$ symmetries. The left--moving
$\{y,\omega\}$ states are also divided into the sets $\{{y}^{3,\cdots,6}\}$, $\{{y}^{1,2}
,{\omega}^{5,6}\}$, $\{{\omega}^{1,\cdots,4}\}$. The left--moving
$\chi^{12},\chi^{34},\chi^{56}$ states carry the supersymmetry charges.
Each sector $\mb_1$, $\mb_2$ and $\mb_3$ carries periodic boundary conditions
under $(\psi^\mu\vert{\bar\psi}^{1,\cdots,5})$ and one of the three groups:
$(\chi_{12},\{y^{3,\cdots,6}\vert{\bar y}^{3,\cdots6}\},{\bar\eta}^1)$,
$(\chi_{34},\{y^{1,2},\omega^{5,6}\vert{\bar y}^{1,2}{\bar\omega}^{5,6}\},
{\bar\eta}^2)$, 
$(\chi_{56},\{\omega^{1,\cdots,4}\vert{\bar\omega}^{1,
\cdots4}\},{\bar\eta}^3)$. 

The second stage of the basis construction consist of adding three
additional basis vectors to the NAHE set. 
Three additional vectors are needed to reduce the number of generations 
to three, one from each sector $\mb_1$, $\mb_2$ and $\mb_3$. 
One specific example is given in Table (\ref{model1}). 
The choice of boundary
conditions to the set of real internal fermions
${\{y,\omega\vert{\bar y},{\bar\omega}\}^{1,\cdots,6}}$  
determines the low energy properties, such as the number of generations,
Higgs doublet--triplet splitting and Yukawa couplings. 

The $SO(10)$ gauge group is broken to one of its subgroups
$SU(5)\times U(1)$, $SO(6)\times SO(4)$ or
$SU(3)\times SU(2)\times U(1)^2$ by the assignment of
boundary conditions to the set ${\bar\psi}^{1\cdots5}_{1\over2}$:

1. $b\{{{\bar\psi}_{1\over2}^{1\cdots5}}\}=
\{{1\over2}{1\over2}{1\over2}{1\over2}
{1\over2}\}\Rightarrow SU(5)\times U(1)$,

2. $b\{{{\bar\psi}_{1\over2}^{1\cdots5}}\}=\{1 1 1 0 0\}
  \Rightarrow SO(6)\times SO(4)$.

To break the $SO(10)$ symmetry to
$SU(3)_C\times SU(2)_L\times
U(1)_C\times U(1)_L$\footnote{$U(1)_C={3\over2}U(1)_{B-L};
U(1)_L=2U(1)_{T_{3_R}}.$}
both steps, 1 and 2, are used, in two separate basis vectors.
Similarly, the breaking pattern
$SO(10)\rightarrow SU(3)_C\times SU(2)_L\times SU(2)_R \times U(1)_{B-L}$
is achieved by the following assignment in two separate basis 
vectors 

1. $b\{{{\bar\psi}_{1\over2}^{1\cdots5}}\}=\{1 1 1 0 0\}
  \Rightarrow SO(6)\times SO(4)$,

2. $b\{{{\bar\psi}_{1\over2}^{1\cdots5}}\}=
\{{1\over2}{1\over2}{1\over2}00\}\Rightarrow SU(3)_C\times U(1)_C
\times SU(2)_L\times SU(2)_R$.

We comment here that 
a recurring feature of some of the three generation free fermionic heterotic
string models is the emergence of a combination of the basis vectors
which extend the NAHE set,
\begin{equation}
\mX=n_\malpha\malpha+n_\mbeta\mbeta+n_\mgamma\mgamma\label{xcomb}
\end{equation}
for which $\mX_L\cdot \mX_L=0$ and $\mX_R\cdot \mX_R\ne0$. Such a 
combination may produce additional space--time vector 
bosons, depending on the choice of GSO phases.
These additional space--time vector bosons
enhance the four dimensional gauge group.
This situation is similar to the presence of the combination
of the NAHE set basis vectors ${\mathbf 1}+\mb_1+\mb_2+\mb_3$, which
enhances the hidden gauge group, at the level of the NAHE set,
{}from $SO(16)$ to $E_8$.
As we discuss below, we often find, although not always, that either the 
$SU(3)_C$ or the $U(1)_C$ symmetry is enhanced to $SU(4)_C$ or
$SU(2)_C$, respectively.
Therefore, we will present models with and without gauge symmetry 
enhancement. In the free fermionic models this type of
gauge symmetry enhancement in the observable sector is,
in general, family universal and is intimately related to the
$\IZ_2\times \IZ_2$ orbifold structure which underlies the
realistic free fermionic models. Such 
enhanced symmetries were shown to forbid proton decay
mediating operators to all orders of nonrenormalizable terms \cite{custodial}.

\section{Left--right symmetric models without enhanced symmetry}\label{noes}

As our first example of a left--right symmetric
free fermionic heterotic string model
we consider Model 1, specified below.
The boundary conditions of the three basis vectors
which extend the NAHE set are shown in Table (\ref{model1}).
Also given in Table (\ref{model1}) are the pairings of 
left-- and right--moving real fermions from the set
$\{y,\omega|{\bar y},{\bar\omega}\}$. These fermions are
paired to form either complex, left-- or right--moving, fermions, 
or Ising model operators, which combine a real left--moving fermion with
a real right--moving fermion. The generalized GSO coefficients
determining the physical massless states of Model 1 appear in 
matrix (\ref{phasesmodel1}).
\vskip 0.4truecm

\LRS Model 1 Boundary Conditions:
\beqn
 &\begin{tabular}{c|c|ccc|c|ccc|c}
 ~ & $\psi^\mu$ & $\chi^{12}$ & $\chi^{34}$ & $\chi^{56}$ &
        $\bar{\psi}^{1,...,5} $ &
        $\bar{\eta}^1 $&
        $\bar{\eta}^2 $&
        $\bar{\eta}^3 $&
        $\bar{\phi}^{1,...,8} $ \\
\hline
\hline
  ${\malpha}$  &  0 & 0&0&0 & 1~1~1~0~0 & 0 & 0 & 0 &1~1~1~1~0~0~0~0 \\
  ${\mbeta}$   &  0 & 0&0&0 & 1~1~1~0~0 & 0 & 0 & 0 &1~1~0~0~1~1~0~0 \\
  ${\mgamma}$  &  0 & 0&0&0 &
${1\over2}$~${1\over2}$~${1\over2}$~0~0&${1\over2}$&${1\over2}$&${1\over2}$ &
\end{tabular}
   \nonumber\\
   ~  &  ~ \nonumber\\
   ~  &  ~ \nonumber\\
     &\begin{tabular}{c|c|c|c}
 ~&   $y^3{y}^6$
      $y^4{\bar y}^4$
      $y^5{\bar y}^5$
      ${\bar y}^3{\bar y}^6$
  &   $y^1{\omega}^5$
      $y^2{\bar y}^2$
      $\omega^6{\bar\omega}^6$
      ${\bar y}^1{\bar\omega}^5$
  &   $\omega^2{\omega}^4$
      $\omega^1{\bar\omega}^1$
      $\omega^3{\bar\omega}^3$
      ${\bar\omega}^2{\bar\omega}^4$ \\
\hline
\hline
$\malpha$& 1 ~~~ 0 ~~~ 0 ~~~ 0  & 0 ~~~ 0 ~~~ 1 ~~~ 1  & 0 ~~~ 0 ~~~ 1 ~~~ 1 \\
$\mbeta$ & 0 ~~~ 0 ~~~ 1 ~~~ 1  & 1 ~~~ 0 ~~~ 0 ~~~ 0  & 0 ~~~ 1 ~~~ 0 ~~~ 1 \\
$\mgamma$& 0 ~~~ 0 ~~~ 1 ~~~ 0  & 1 ~~~ 0 ~~~ 0 ~~~ 1  & 0 ~~~ 1 ~~~ 0 ~~~ 0 \\
\end{tabular}
\label{model1}
\eeqn

\LRS Model 1 Generalized GSO Coefficients:
\begin{equation}
{\bordermatrix{
      &{\bf 1}&\mS & & {\mb_1}&{\mb_2}&{\mb_3}& & {\malpha}&{\mbeta}&{\mgamma}\cr
       {\bf 1}&~~1&~~1 & & -1   &  -1 & -1  & & ~~1     & ~~1   & ~~i   \cr
           \mS&~~1&~~1 & &~~1   & ~~1 &~~1  & &  -1     &  -1   &  -1   \cr
	      &   &    & &      &     &     & &         &       &       \cr
       {\mb_1}& -1& -1 & & -1   &  -1 & -1  & &  -1     &  -1   & ~~1   \cr
       {\mb_2}& -1& -1 & & -1   &  -1 & -1  & &  -1     &  -1   & ~~1   \cr
       {\mb_3}& -1& -1 & & -1   &  -1 & -1  & &  -1     & ~~1   & ~~1   \cr
	      &   &    & &      &     &     & &         &       &       \cr
     {\malpha}&~~1& -1 & &~~1   &  -1 & -1  & &  -1     & ~~1   & ~~i   \cr
     { \mbeta}&~~1& -1 & & -1   & ~~1 &~~1  & &  -1     &  -1   & ~~i   \cr
     {\mgamma}&~~1& -1 & & -1   &  -1 & -1  & &  -1     & ~~1   & ~~1   \cr}}
\label{phasesmodel1}
\end{equation}

In matrix (\ref{phasesmodel1}) only the entries above the diagonal are
independent and those below and on the diagonal are fixed by
the modular invariance constraints. 
Blank lines are inserted to emphasize the division of the free
phases between the different sectors of the realistic
free fermionic models. Thus, the first two lines involve
only the GSO phases of $c{{\{{\bf 1},\mS\}}\choose \ma_i}$. The set
$\{{\bf 1},\mS\}$ generates the $N=4$ model with $\mS$ being the
space--time supersymmetry generator and therefore the phases
$c{\mS\choose{\ma_i}}$ are those that control the space--time supersymmetry
in the superstring models. Similarly, in the free fermionic
models, sectors with periodic and anti--periodic boundary conditions,
of the form of $\mb_i$, produce the chiral generations.
The phases $c{\mb_i\choose \mb_j}$ determine the chirality
of the states from these sectors. 

In the free fermionic models
the basis vectors $\mb_i$ are those that respect the $SO(10)$ symmetry
while the vectors denoted by Greek letters are those that break the
$SO(10)$ symmetry.
As the Standard Model matter states arise from sectors which
preserve the $SO(10)$ symmetry, the phases that fix
the Standard Model charges are, in general, 
the phases $c{\mb_i\choose{\ma_i}}$. On the other hand,
the basis vectors of the form $\{\malpha,\mbeta,\mgamma\}$ break the
$SO(10)$ symmetry. The phases associated with these basis vectors
are associated with exotic physics, beyond the Standard Model. 
These phases, therefore, also affect the final four dimensional
gauge symmetry.

The final gauge group in Model 1 arises
as follows: In the observable sector the NS boundary conditions 
produce gauge group generators for 
\beq
SU(3)_C\times SU(2)_L\times SU(2)_R\times U(1)_C\times U(1)_{1,2,3}\times
U(1)_{4,5,6}
\eeq
Thus, the $SO(10)$ symmetry is broken to
$SU(3)\times SU(2)_L\times SU(2)_R\times U(1)_C$, as discussed above,
where, 
\begin{equation}
U(1)_C={\rm Tr}\, U(3)_C~\Rightarrow~Q_C=
			 \sum_{i=1}^3Q({\bar\psi}^i).
\label{u1c}
\end{equation}
The flavor $SO(6)^3$ symmetries are broken to $U(1)^{3+n}$ with
$(n=0,\cdots,6)$. The first three, denoted by $U(1)_{j}$ $(j=1,2,3)$, arise 
{}from the world--sheet currents ${\bar\eta}^j{\bar\eta}^{j^*}$.
These three $U(1)$ symmetries are present in all
the three generation free fermionic models which use the NAHE set. 
Additional horizontal $U(1)$ symmetries, denoted by $U(1)_{j}$ 
$(j=4,5,...)$, arise by pairing two real fermions from the sets
$\{{\bar y}^{3,\cdots,6}\}$, 
$\{{\bar y}^{1,2},{\bar\omega}^{5,6}\}$, and
$\{{\bar\omega}^{1,\cdots,4}\}$. 
The final observable gauge group depends on
the number of such pairings. In this model there are the 
pairings, ${\bar y}^3{\bar y}^6$, ${\bar y}^1{\bar\omega}^5$
and ${\bar\omega}^2{\bar\omega}^4$, which generate three additional 
$U(1)$ symmetries, denoted by $U(1)_{4,5,6}$\footnote{It is 
important to note that the existence of these three additional 
$U(1)$ currents is correlated with a superstringy doublet--triplet
splitting mechanism \cite{ps}. Due to these extra $U(1)$ symmetries 
the color triplets from the NS sector are projected out of the spectrum 
by the GSO projections while the electroweak doublets remain in the 
light spectrum.}. 

In the hidden sector the NS boundary conditions produce the generators of
\beq
SU(2)_1\times U(1)_{H_1}\times SU(2)_2\times U(1)_{H_2}\times 
U(1)_{7,8,9,10}
\label{nshiden}
\eeq
where $SU(2)_1$ and $SU(2)_2$ arise from the complex world--sheet fermions
$\{{\bar\phi}^3,{\bar\phi}^4\}$ and $\{{\bar\phi}^5,{\bar\phi}^6\}$,
respectively; and $U(1)_{H_1}$ and $U(1)_{H_2}$
correspond to the combinations of world--sheet charges
\begin{eqnarray}
Q_{H_1}&=&Q({\bar\phi^1})-Q({\bar\phi^2}) +
\sum_{i=5}^7Q({\bar\phi})^i-Q({\bar\phi})^8,\label{qh1model1}\\
Q_{H_2}&=&\sum_{i=1}^4Q({\bar\phi})^i -Q({\bar\phi^7}) - Q({\bar\phi})^8 .
\label{qh2model1}
\end{eqnarray}
The charges under the 
remaining four orthogonal $U(1)$ combinations are given by
\begin{eqnarray}
Q_{7}&=&~~Q({\bar\phi}^1)+Q({\bar\phi}^8),\nonumber\\
Q_{8}&=&~~Q({\bar\phi}^2)+Q({\bar\phi}^7),\nonumber\\
Q_{9}&=&~~Q({\bar\phi}^1)-Q({\bar\phi}^3)-Q({\bar\phi}^4)-
Q({\bar\phi}^5)-Q({\bar\phi}^6)-Q({\bar\phi}^8),\nonumber\\
Q_{10}&=&~~Q({\bar\phi}^2)-Q({\bar\phi}^3)-Q({\bar\phi}^4)+
Q({\bar\phi}^5)+Q({\bar\phi}^6)-Q({\bar\phi}^7).
\label{u1commodel1}
\end{eqnarray}
The sector $\mzeta\equiv1+\mb_1+\mb_2+\mb_3$ produces the 
representations $(2,1)_{\pm 4,0}$ and $(1,2)_{0,\pm 4}$ of 
$SU(2)_{H_1}\times U(1)_{H_1}$ and $SU(2)_{H_2}\times U(1)_{H_2}$,
raising the symmetry to $SU(3)_{H_1}\times SU(3)_{H_2}$.
Thus, the hidden $E_8$ symmetry is broken to 
$SU(3)_{H_1}\times SU(3)_{H_2}\times U(1)_{7,8,9,10}$. 

In addition to the graviton, dilaton, antisymmetric sector and spin--1 gauge bosons, 
the NS sector gives two pairs 
of electroweak doublets, transforming as (1,2,2,0) under 
$SU(3)_C\times SU(2)_L\times SU(2)_R\times U(1)_C$; 
three pairs of $SO(10)$ singlets with 
$U(1)_{1,2,3}$ charges; and three singlets of the entire four 
dimensional gauge group. 
The states from the sectors $\mb_j$ $(j=1,2,3)$ produce
the three light generations. These states and their
decomposition under the entire gauge group are shown in Table 1 in 
Appendix A. The remaining massless states and their quantum numbers  
also appear in Table 1. 

\subsection{Model 1 Superpotential}

We now turn to the superpotential of the model. The cubic
level and higher order terms in the superpotential are 
obtained by calculating the correlators between the vertex
operators. The non--vanishing terms must be invariant under
all the symmetries of the string models and must satisfy
all the string selection rules \cite{kln}.
The full superpotential has been analyzed up to order $N=6$. 
Below we give the cubic and quartic order terms and the 
quintic order terms are given in Appendix B. 
We divide the superpotential terms into four sets. 
Terms in the first set contain the states 
that transform nontrivialy under the Standard Model gauge group. 
Terms in the second set contain only states that are singlets 
of all non--Abelian groups.
Terms in the third set contain states that transform nontrivialy under the
unbroken hidden $E_8$ non--Abelian subgroup, while terms in the fourth set
contain both Standard Model and Hidden Sector states.
We indicate when no terms of a given type are found at a specific
order. 
\newpage

{
\textwidth=7.5in
\oddsidemargin=-18mm
\renewcommand{\baselinestretch}{1.3}

\begin{flushleft}
\no $W_3({\rm observable})$:
\beqn 
{\begin{tabular}{llllll}
   $\php \h{1 }  \h{2 }  \P{3 } $  
 & $+ \h{1 }  \Q{L_1}  \Q{R_1} $  
 & $+ \h{1 }  \L{L_1}  \L{R_1} $  
 & $+ \h{1 } \cL{+}{L_{23}} \cL{-}{R_{23}} $  
 & $+ \h{1 } \cL{-}{L_{23}} \cL{+}{R_{23}} $  
 & $+ \h{1 } \cL{+}{L_1} \cL{-}{R_1} $ \\
   $+ \h{1 } \cL{-}{L_1} \cL{+}{R_1} $  
 & $+ \h{2 }  \Q{L_2}  \Q{R_2} $  
 & $+ \h{2 }  \L{L_2}  \L{R_2} $  
 & $+ \h{2 } \cL{+}{L_{13}} \cL{-}{R_{13}} $  
 & $+ \h{2 } \cL{-}{L_{13}} \cL{+}{R_{13}} $  
 & $+ \h{2 } \cL{+}{L_2} \cL{-}{R_2} $ \\
   $+ \h{2 } \cL{-}{L_2} \cL{+}{R_2} $  
 & $+ \Q{L_3} \bD{\ab} \cL{-}{L_3} $  
 & $+ \Q{R_3}  \D{\ab} \cL{+}{R_3} $  
 & $+ \L{L_1} \cL{+}{L_1} \bx{5 } $  
 & $+ \L{L_2} \cL{+}{L_2} \bx{6 } $  
 & $+ \L{L_3} \cL{+}{L_3} \bx{7 } $ \\
   $+ \L{R_1} \cL{-}{R_1}  \x{5 } $  
 & $+ \L{R_2} \cL{-}{R_2}  \x{6 } $  
 & $+ \L{R_3} \cL{-}{R_3}  \x{7 } $  
 & $+\cL{-}{L_{13}} \cL{-}{L_2} \bx{9 } $  
 & $+\cL{+}{L_{23}} \cL{-}{L_1} \bx{8 } $  
 & $+\cL{+}{R_{13}} \cL{+}{R_2}  \x{9 } $ \\
   $+\cL{-}{R_{23}} \cL{+}{R_1}  \x{8 } $  
\end{tabular}} 
\label{w3ssm2a} 
\eeqn

\no $W_3({\rm singlets})$:
\beqn 
{\begin{tabular}{llllll}
   $\php \P{3 }  \x{1 } \bx{1 } $  
 & $+ \P{3 }  \x{2 } \bx{2 } $  
 & $+ \P{3 }  \x{3 } \bx{3 } $  
 & $+ \P{3 }  \x{4 } \bx{4 } $  
 & $+ \P{3 }  \x{10} \bx{10} $  
 & $+ \P{3 }  \x{11} \bx{11} \phantom{+ \P{3 }  \H{2 } \bH{2 }} $ \\
   $+ \P{12 } \bP{13}  \P{23} $  
 & $+ \P{12 }  \x{1 }  \x{4 } $  
 & $+ \P{12 }  \x{2 }  \x{3 } $  
 & $+ \P{12 }  \x{6 } \bx{5 } $  
 & $+\bP{12 }  \P{13} \bP{23} $  
 & $+\bP{12 } \bx{1 } \bx{4 } \phantom{+ \P{3 }  \H{2 } \bH{2 }} $ \\
   $+\bP{12 } \bx{2 } \bx{3 } $  
 & $+\bP{12 } \bx{6 }  \x{5 } $  
 & $+ \P{13}  \x{7 } \bx{5 } $  
 & $+\bP{13} \bx{7 }  \x{5 } $  
 & $+ \P{23}  \x{7 } \bx{6 } $  
 & $+\bP{23} \bx{7 }  \x{6 } \phantom{+ \P{3 }  \H{2 } \bH{2 }} $ \\
\end{tabular}} 
\label{w3sig2a} 
\eeqn

\no $W_3({\rm hidden})$:
\beqn 
{\begin{tabular}{llllll}
   $\php  \P{3 }  \H{1 } \bH{1 } $  
 & $+ \P{3 }  \H{2 } \bH{2 } $  
 & $\phantom{+\P{3 }  \x{4 } \bx{4 }} $  
 & $\phantom{+\P{3 }  \x{4 } \bx{4 }} $  
 & $\phantom{+\P{3 }  \x{10} \bx{10}} $  
 & $\phantom{+\P{3 }  \x{11} \bx{11} + \P{3 }  \H{2 } \bH{2 }} $ 
\end{tabular}} 
\label{w3hid2a} 
\eeqn

\no $W_4({\rm observable})$:
\beqn 
{\begin{tabular}{lllll}
   $\php  \Q{L_1}  \Q{L_3}  \Q{R_1}  \Q{R_3} $  
 & $+ \Q{L_1}  \Q{R_1}  \L{L_3}  \L{R_3} $  
 & $+ \Q{L_1}  \Q{R_1} \cL{+}{L_{12}} \cL{-}{R_{12}} $  
 & $+ \Q{L_1}  \Q{R_1} \cL{-}{L_{12}} \cL{+}{R_{12}} $  
 & $+ \Q{L_1}  \Q{R_1} \cL{+}{L_3} \cL{-}{R_3} $ \\
   $+ \Q{L_1}  \Q{R_1} \cL{-}{L_3} \cL{+}{R_3} $  
 & $+ \Q{L_1}  \Q{R_2} \cL{+}{L_{13}} \cL{-}{R_{23}} $  
 & $+ \Q{L_1}  \Q{R_2} \cL{-}{L_{13}} \cL{+}{R_{23}} $  
 & $+ \Q{L_1}  \Q{R_3} \cL{+}{L_{12}} \cL{-}{R_{23}} $  
 & $+ \Q{L_2}  \Q{L_3}  \Q{R_2}  \Q{R_3} $ \\
   $+ \Q{L_2}  \Q{R_1} \cL{+}{L_{23}} \cL{-}{R_{13}} $  
 & $+ \Q{L_2}  \Q{R_1} \cL{-}{L_{23}} \cL{+}{R_{13}} $  
 & $+ \Q{L_2}  \Q{R_2}  \L{L_3}  \L{R_3} $  
 & $+ \Q{L_2}  \Q{R_2} \cL{+}{L_{12}} \cL{-}{R_{12}} $  
 & $+ \Q{L_2}  \Q{R_2} \cL{-}{L_{12}} \cL{+}{R_{12}} $ \\
   $+ \Q{L_2}  \Q{R_2} \cL{+}{L_3} \cL{-}{R_3} $  
 & $+ \Q{L_2}  \Q{R_2} \cL{-}{L_3} \cL{+}{R_3} $  
 & $+ \Q{L_2}  \Q{R_3} \cL{-}{L_{12}} \cL{+}{R_{13}} $  
 & $+ \Q{L_3}  \Q{R_1} \cL{+}{L_{23}} \cL{-}{R_{12}} $  
 & $+ \Q{L_3}  \Q{R_2} \cL{-}{L_{13}} \cL{+}{R_{12}} $ \\
   $+ \Q{L_3}  \Q{R_3}  \L{L_1}  \L{R_1} $  
 & $+ \Q{L_3}  \Q{R_3}  \L{L_2}  \L{R_2} $  
 & $+ \Q{L_3}  \Q{R_3} \cL{+}{L_{13}} \cL{-}{R_{13}} $  
 & $+ \Q{L_3}  \Q{R_3} \cL{-}{L_{13}} \cL{+}{R_{13}} $  
 & $+ \Q{L_3}  \Q{R_3} \cL{+}{L_{23}} \cL{-}{R_{23}} $ \\
   $+ \Q{L_3}  \Q{R_3} \cL{-}{L_{23}} \cL{+}{R_{23}} $  
 & $+ \Q{L_3}  \Q{R_3} \cL{+}{L_1} \cL{-}{R_1} $  
 & $+ \Q{L_3}  \Q{R_3} \cL{-}{L_1} \cL{+}{R_1} $  
 & $+ \Q{L_3}  \Q{R_3} \cL{+}{L_2} \cL{-}{R_2} $  
 & $+ \Q{L_3}  \Q{R_3} \cL{-}{L_2} \cL{+}{R_2} $ \\
   $+ \L{L_1}  \L{L_3}  \L{R_1}  \L{R_3} $  
 & $+ \L{L_1}  \L{R_1} \cL{+}{L_{12}} \cL{-}{R_{12}} $  
 & $+ \L{L_1}  \L{R_1} \cL{-}{L_{12}} \cL{+}{R_{12}} $  
 & $+ \L{L_1}  \L{R_1} \cL{+}{L_3} \cL{-}{R_3} $  
 & $+ \L{L_1}  \L{R_1} \cL{-}{L_3} \cL{+}{R_3} $ \\
   $+ \L{L_1} \cL{-}{L_{23}}  \x{1 }  \x{11} $  
 & $+ \L{L_1} \cL{-}{L_{23}}  \x{2 }  \x{10} $  
 & $+ \L{L_2}  \L{L_3}  \L{R_2}  \L{R_3} $  
 & $+ \L{L_2}  \L{R_2} \cL{+}{L_{12}} \cL{-}{R_{12}} $  
 & $+ \L{L_2}  \L{R_2} \cL{-}{L_{12}} \cL{+}{R_{12}} $ \\
   $+ \L{L_2}  \L{R_2} \cL{+}{L_3} \cL{-}{R_3} $  
 & $+ \L{L_2}  \L{R_2} \cL{-}{L_3} \cL{+}{R_3} $  
 & $+ \L{L_2} \cL{+}{L_{13}} \bx{1 }  \x{10} $  
 & $+ \L{L_2} \cL{+}{L_{13}} \bx{2 }  \x{11} $  
 & $+ \L{L_3}  \L{R_3} \cL{+}{L_{13}} \cL{-}{R_{13}} $ \\
   $+ \L{L_3}  \L{R_3} \cL{-}{L_{13}} \cL{+}{R_{13}} $  
 & $+ \L{L_3}  \L{R_3} \cL{+}{L_{23}} \cL{-}{R_{23}} $  
 & $+ \L{L_3}  \L{R_3} \cL{-}{L_{23}} \cL{+}{R_{23}} $  
 & $+ \L{L_3}  \L{R_3} \cL{+}{L_1} \cL{-}{R_1} $  
 & $+ \L{L_3}  \L{R_3} \cL{-}{L_1} \cL{+}{R_1} $ \\
   $+ \L{L_3}  \L{R_3} \cL{+}{L_2} \cL{-}{R_2} $  
 & $+ \L{L_3}  \L{R_3} \cL{-}{L_2} \cL{+}{R_2} $  
 & $+ \L{R_1} \cL{+}{R_{23}} \bx{1 } \bx{11} $  
 & $+ \L{R_1} \cL{+}{R_{23}} \bx{2 } \bx{10} $  
 & $+ \L{R_2} \cL{-}{R_{13}}  \x{1 } \bx{10} $ \\
   $+ \L{R_2} \cL{-}{R_{13}}  \x{2 } \bx{11} $  
 & $+\cL{+}{L_{12}} \cL{+}{L_{13}} \cL{-}{R_{12}} \cL{-}{R_{13}} $  
 & $+\cL{+}{L_{12}} \cL{-}{L_{13}} \cL{-}{R_{12}} \cL{+}{R_{13}} $  
 & $+\cL{+}{L_{12}} \cL{+}{L_{23}} \cL{-}{R_{12}} \cL{-}{R_{23}} $  
 & $+\cL{+}{L_{12}} \cL{-}{L_{23}} \cL{-}{R_{12}} \cL{+}{R_{23}} $ \\
   $+\cL{+}{L_{12}} \cL{+}{L_1} \cL{-}{R_{12}} \cL{-}{R_1} $  
 & $+\cL{+}{L_{12}} \cL{-}{L_1} \cL{-}{R_{12}} \cL{+}{R_1} $  
 & $+\cL{+}{L_{12}} \cL{+}{L_2} \cL{-}{R_{12}} \cL{-}{R_2} $  
 & $+\cL{+}{L_{12}} \cL{+}{L_2} \cL{-}{R_{13}} \cL{-}{R_3} $  
 & $+\cL{+}{L_{12}} \cL{-}{L_2} \cL{-}{R_{12}} \cL{+}{R_2} $ \\
   $+\cL{-}{L_{12}} \cL{+}{L_{13}} \cL{+}{R_{12}} \cL{-}{R_{13}} $  
 & $+\cL{-}{L_{12}} \cL{-}{L_{13}} \cL{+}{R_{12}} \cL{+}{R_{13}} $  
 & $+\cL{-}{L_{12}} \cL{+}{L_{23}} \cL{+}{R_{12}} \cL{-}{R_{23}} $  
 & $+\cL{-}{L_{12}} \cL{-}{L_{23}} \cL{+}{R_{12}} \cL{+}{R_{23}} $  
 & $+\cL{-}{L_{12}} \cL{+}{L_1} \cL{+}{R_{12}} \cL{-}{R_1} $ \\
   $+\cL{-}{L_{12}} \cL{+}{L_1} \cL{+}{R_{23}} \cL{-}{R_3} $  
 & $+\cL{-}{L_{12}} \cL{-}{L_1} \cL{+}{R_{12}} \cL{+}{R_1} $  
 & $+\cL{-}{L_{12}} \cL{+}{L_2} \cL{+}{R_{12}} \cL{-}{R_2} $  
 & $+\cL{-}{L_{12}} \cL{-}{L_2} \cL{+}{R_{12}} \cL{+}{R_2} $  
 & $+\cL{+}{L_{13}} \cL{+}{L_1} \cL{-}{R_{23}} \cL{-}{R_2} $ \\
   $+\cL{+}{L_{13}} \cL{+}{L_3} \cL{-}{R_{12}} \cL{-}{R_2} $  
 & $+\cL{+}{L_{13}} \cL{+}{L_3} \cL{-}{R_{13}} \cL{-}{R_3} $  
 & $+\cL{+}{L_{13}} \cL{-}{L_3} \cL{-}{R_{13}} \cL{+}{R_3} $  
 & $+\cL{-}{L_{13}} \cL{+}{L_1} \cL{+}{R_{23}} \cL{-}{R_2} $  
 & $+\cL{-}{L_{13}} \cL{+}{L_3} \cL{+}{R_{13}} \cL{-}{R_3} $ \\
   $+\cL{-}{L_{13}} \cL{-}{L_3} \cL{+}{R_{13}} \cL{+}{R_3} $  
 & $+\cL{+}{L_{23}} \cL{+}{L_2} \cL{-}{R_{13}} \cL{-}{R_1} $  
 & $+\cL{+}{L_{23}} \cL{+}{L_3} \cL{-}{R_{23}} \cL{-}{R_3} $  
 & $+\cL{+}{L_{23}} \cL{-}{L_3} \cL{-}{R_{23}} \cL{+}{R_3} $  
 & $+\cL{-}{L_{23}} \cL{+}{L_2} \cL{+}{R_{13}} \cL{-}{R_1} $ \\
   $+\cL{-}{L_{23}} \cL{+}{L_3} \cL{+}{R_{12}} \cL{-}{R_1} $  
 & $+\cL{-}{L_{23}} \cL{+}{L_3} \cL{+}{R_{23}} \cL{-}{R_3} $  
 & $+\cL{-}{L_{23}} \cL{-}{L_3} \cL{+}{R_{23}} \cL{+}{R_3} $  
 & $+\cL{+}{L_1} \cL{+}{L_3} \cL{-}{R_1} \cL{-}{R_3} $  
 & $+\cL{+}{L_1} \cL{-}{L_3} \cL{-}{R_1} \cL{+}{R_3} $ \\
   $+\cL{-}{L_1} \cL{+}{L_3} \cL{+}{R_1} \cL{-}{R_3} $  
 & $+\cL{-}{L_1} \cL{-}{L_3} \cL{+}{R_1} \cL{+}{R_3} $  
 & $+\cL{+}{L_2} \cL{+}{L_3} \cL{-}{R_2} \cL{-}{R_3} $  
 & $+\cL{+}{L_2} \cL{-}{L_3} \cL{-}{R_2} \cL{+}{R_3} $  
 & $+\cL{-}{L_2} \cL{+}{L_3} \cL{+}{R_2} \cL{-}{R_3} $ \\
   $+\cL{-}{L_2} \cL{-}{L_3} \cL{+}{R_2} \cL{+}{R_3} $  
\end{tabular}} 
\label{w4ssm2a} 
\eeqn

\no $W_4({\rm singlets})$, $W_4({\rm mixed})$, $W_4({\rm hidden})$: none
\end{flushleft}
\newpage
}

\section{Models with enhanced non--Abelian symmetries}\label{es}

We next turn to our second example, Model 2. The boundary condition basis 
vectors and one--loop phases, which define the model, are given in 
Table~(\ref{model2}) and matrix (\ref{phasesmodel2}), respectively. 
\vskip 0.4truecm

\LRS Model 2 Boundary Conditions:
\beqn
 &\begin{tabular}{c|c|ccc|c|ccc|c}
 ~ & $\psi^\mu$ & $\chi^{12}$ & $\chi^{34}$ & $\chi^{56}$ &
        $\bar{\psi}^{1,...,5} $ &
        $\bar{\eta}^1 $&
        $\bar{\eta}^2 $&
        $\bar{\eta}^3 $&
        $\bar{\phi}^{1,...,8} $ \\
\hline
\hline
 ${\malpha}$  &  0 & 0&0&0 & 1~1~1~0~0 & 0 & 0 & 0 &1~1~1~1~0~0~0~0 \\
 ${\mbeta}$   &  0 & 0&0&0 & 1~1~1~0~0 & 0 & 0 & 0 &1~1~1~1~0~0~0~0 \\
 ${\mgamma}$  &  0 & 0&0&0 &
		${1\over2}$~${1\over2}$~${1\over2}$~0~0
		& ${1\over2}$ & ${1\over2}$ & ${1\over2}$ &
\end{tabular}
   \nonumber\\
   ~  &  ~ \nonumber\\
   ~  &  ~ \nonumber\\
     &\begin{tabular}{c|c|c|c}
 ~&   $y^3{y}^6$
      $y^4{\bar y}^4$
      $y^5{\bar y}^5$
      ${\bar y}^3{\bar y}^6$
  &   $y^1{\omega}^5$
      $y^2{\bar y}^2$
      $\omega^6{\bar\omega}^6$
      ${\bar y}^1{\bar\omega}^5$
  &   $\omega^2{\omega}^4$
      $\omega^1{\bar\omega}^1$
      $\omega^3{\bar\omega}^3$
      ${\bar\omega}^2{\bar\omega}^4$ \\
\hline
\hline
$\malpha$& 1 ~~~ 0 ~~~ 0 ~~~ 0  & 0 ~~~ 0 ~~~ 1 ~~~ 1  & 0 ~~~ 0 ~~~ 1 ~~~ 1 \\
$\mbeta$ & 0 ~~~ 0 ~~~ 1 ~~~ 1  & 1 ~~~ 0 ~~~ 0 ~~~ 0  & 0 ~~~ 1 ~~~ 0 ~~~ 1 \\
$\mgamma$& 0 ~~~ 0 ~~~ 1 ~~~ 0  & 1 ~~~ 0 ~~~ 0 ~~~ 1  & 0 ~~~ 1 ~~~ 0 ~~~ 0 \\
\end{tabular}
\label{model2}
\eeqn

\LRS Model 2 Generalized GSO Coefficients:
\begin{equation}
{\bordermatrix{
              &{\bf 1}&\mS & & {\mb_1}&{\mb_2}&{\mb_3}& &{\malpha}&{\mbeta}&{\mgamma}\cr
       {\bf 1}&~~1&~~1 & & -1   &  -1 & -1  & & ~~1     & ~~1   & ~~i   \cr
           \mS&~~1&~~1 & &~~1   & ~~1 &~~1  & &  -1     &  -1   &  -1   \cr
	      &   &    & &      &     &     & &         &       &       \cr
       {\mb_1}& -1& -1 & & -1   &  -1 & -1  & &  -1     &  -1   & ~~1   \cr
       {\mb_2}& -1& -1 & & -1   &  -1 & -1  & &  -1     &  -1   & ~~1   \cr
       {\mb_3}& -1& -1 & & -1   &  -1 & -1  & &  -1     &  -1   & ~~1   \cr
	      &   &    & &      &     &     & &         &       &       \cr
     {\malpha}&~~1& -1 & &~~1   &  -1 & -1  & &  -1     & ~~1   & ~~i   \cr
     {\mbeta} &~~1& -1 & & -1   & ~~1 & -1  & & ~~1     &  -1   & ~~i   \cr
     {\mgamma}&~~1& -1 & & -1   &  -1 & -1  & &  -1     & ~~1   & ~~1   \cr}}
\label{phasesmodel2}
\end{equation}

Model 2, defined by Table (\ref{model2}) and matrix (\ref{phasesmodel2}), 
differs from Model 1 only in the boundary conditions
of the hidden sector world--sheet fermions 
$\{{\bar\phi}^{1,\cdots,8}\}$ in the basis vectors $\mbeta$ and $\mgamma$, 
and in the GSO phases beyond the NAHE ones. In the basis vector
$\mbeta$ the alterations are 
$\beta_{{\bar\phi}^{3,4}}= 0\rightarrow1$;
$\beta_{{\bar\phi}^{5,6}}=1\rightarrow0$ and in $\mgamma$ they are
$\gamma_{{\bar\phi}^1}=1\rightarrow{1\over2}$;
$\gamma_{{\bar\phi}^4}={1\over2}\rightarrow1$.
While the spectrum arising from the NAHE 
set remains essentially unaltered, the spectrum  
arising from the basis vectors beyond the NAHE set 
is substantially modified. Hence some phenomenological
features of the two models are significantly modified.  

The total gauge group of Model 2 arises as follows. 
In the observable sector the NS boundary conditions produce the  
generators of 
$(SU(3)_C \times U(1)_C \times SU(2)_L\times SU(2)_R\in SO(10))   
\times U(1)_{1,2,3}\times U(1)_{4,5,6}$,
%
while in the hidden sector 
the NS boundary conditions produce the generators of
\beq
SU(3)_{H_1}\times U(1)_{H_1}\times U(1)_7\times
SU(3)_{H_2}\times U(1)_{H_2}\times U(1)_8\, .
\eeq
$U(1)_{H_1}$ and $U(1)_{H_2}$
correspond to the combinations of the world--sheet charges
\begin{eqnarray}
Q_{H_1}&=&Q({\bar\phi^1})-Q({\bar\phi^2})-Q({\bar\phi^3})+
\sum_{i=4}^7Q({\bar\phi})^i-Q({\bar\phi})^8,\label{qh1}\\
Q_{H_2}&=&\sum_{i=1}^4Q({\bar\phi})^i -Q({\bar\phi^7})
-Q({\bar\phi})^8 .
\label{qh2}
\end{eqnarray}
and $U(1)_{7,8}$ arise from the world--sheet currents 
${\bar\phi}^4{\bar\phi}^{4^*}$ and ${\bar\phi}^8{\bar\phi}^{8^*}$, 
respectively. 
The sector $\mzeta\equiv1+\mb_1+\mb_2+\mb_3$ produces the representations 
$(3,1)_{-5,0}\oplus({\bar3},1)_{5,0}$ and 
$(1,3)_{0,-5}\oplus(1,{\bar3})_{0,5}$
of 
$SU(3)_{H_1}\times U(1)_{H_1}$ and $SU(3)_{H_2}\times U(1)_{H_2}$.
Thus, the $E_8$ symmetry
reduces to $SU(4)_{H_1}\times SU(4)_{H_2}\times U(1)^2$. 
The additional $U(1)$'s in $SU(4)_{H_{1,2}}$ are given by the combinations in
eqs.~(\ref{qh1}) and (\ref{qh2}), respectively.
The remaining $U(1)$ symmetries in the
hidden sector, $U(1)_{7^\prime}$ and $U(1)_{8^\prime}$,
correspond to the combination of world--sheet charges
\begin{eqnarray}
Q_{7^\prime}&=&Q({\bar\phi^4})+Q({\bar\phi^8}),\label{q7p}\\
Q_{8^\prime}&=& -\sum_{i=1}^3Q({\bar\phi})^i
+Q({\bar\phi^4})-\sum_{i=5}^7Q({\bar\phi})^i-Q({\bar\phi})^8 .
\label{q8p}
\end{eqnarray}

Model 2 contains two combinations
of non--NAHE basis vectors with $\mX_L\cdot \mX_L=0$, which
therefore may give rise to additional space--time vector bosons.
The first is the vector combination ${\mbeta\pm\mgamma}$.
The second combination is given by $\mzeta+2\mgamma$, 
where $\mzeta\equiv1+\mb_1+\mb_2+\mb_3$.
The presence of the first combination depends on the 
assignment of periodic boundary conditions in the 
basis vectors $\malpha$, $\mbeta$ and $\mgamma$, which 
extend the NAHE set and is therefore model dependent.
The second combination, however, arises only from the 
NAHE set basis vectors plus $2\mgamma$ and is therefore
independent of the assignment of periodic boundary
conditions in the basis vectors $\malpha$, $\mbeta$ and $\mgamma$.
This vector combination is therefore generic for the 
pattern of symmetry breaking $SO(10)\rightarrow
SU(3)_C\times U(1)_C\times SU(2)_L\times SU(2)_R$,
in NAHE based models. 

In Model 2 all the 
space--time vector bosons from the sector $\mbeta\pm\mgamma$ 
are projected out by the GSO projections and therefore
give no gauge enhancement.
The sector $\mzeta+2\mgamma$, 
however, gives rise to two additional space--time vector bosons 
which are charged with respect to the world--sheet
$U(1)$ currents. This enhances one of the world--sheet
$U(1)$ combinations to $SU(2)_{\rm cust}$. The relevant 
combination of world--sheet charges is given by
\begin{equation}
Q_{SU(2)_C}=Q_C+(Q_{\eta_1}+Q_{\eta_2}+Q_{\eta_3})-Q_{7^\prime}.
\label{u1su2}
\end{equation}
The remaining orthogonal $U(1)$ combinations are 
\begin{eqnarray}
Q_{1^\prime}&=&~~Q_1-Q_2,\nonumber\\
Q_{2^\prime}&=&~~Q_1+Q_2-2Q_3,\nonumber\\
Q_{3^\prime}&=&-Q_C+Q_1+Q_2+Q_3+Q_{8^\prime},\nonumber\\
Q_{7^{\prime\prime}}&=&~~Q_C+Q_1+Q_2+Q_3+3Q_{7^\prime},\nonumber\\
Q_{8^{\prime\prime}}&=&-4Q_C+4(Q_1+Q_2+Q_3)-3Q_{8^\prime}.
\label{u1com}
\end{eqnarray}
and $Q_{4,5,6}$ are unchanged.
Thus, the full massless spectrum transforms under the final gauge group, 
$SU(3)_C\times SU(2)_L\times SU(2)_R\times SU(2)_{\rm cust}\times
U(1)_{1^\prime,2^\prime,3^\prime}\times U(1)_{4,5,6}\times 
SU(4)_{H_1}\times SU(4)_{H_2}\times U(1)_{7^{\prime\prime},8^{\prime\prime}}$. 

In addition to the graviton, 
dilaton, antisymmetric sector and spin--1 gauge bosons, 
the NS sector gives two pairs 
of electroweak doublets, transforming as (1,2,2,1) under 
$SU(3)_C\times SU(2)_L\times SU(2)_R\times SU(2)_{\rm cust}$; 
three pairs of $SO(10)$ singlets with 
$U(1)_{1,2,3}$ charges; and three singlets of the entire four 
dimensional gauge group. The sector $\mS+\mb_1+\mb_2+\malpha+\mbeta$ 
produces one pair of $SU(2)_{\rm cust}$ doublets that can be used to 
break the $SU(2)_{\rm cust}$ symmetry, and three pairs of non--Abelian 
singlets with $U(1)_{1,2,3}$ charges. 

The states from the sectors $\mb_j\oplus \mzeta+2\mgamma~(j=1,2,3)$ produce
the three light generations. The states from these sectors and their
decomposition under the entire gauge group are shown in Table 2. 
The leptons (and quarks) are singlets of the color $SU(4)_{H_1,H_2}$ 
gauge groups and 
the $U(1)_{7^{\prime\prime}}$ symmetry of eq.~(\ref{u1com}) 
becomes a gauged leptophobic symmetry. 
The remaining massless states and their quantum numbers are also 
given in Table 2 in Appendix C. 
We also provide, in Appendix D, the Model 2 cubic through quintic order 
superpotential terms. 

\subsection{Definition of the weak--hypercharge}

We now turn to the definition of the weak--hypercharge 
in this \LRS model. Due to the enhanced
symmetry there are several possibilities to define a weak--hypercharge
combination which is still family universal and reproduces 
the correct charge assignment for the Standard Model fermions. 
As we discuss below, this feature of the free fermionic 
models with enhanced symmetry presents an interesting way to 
understand how the Standard Model spectrum may still arise
{}from $SO(10)$ representation, {\it i.e.}~from the three {\bf 16}'s of
$SO(10)$ which arise from the NAHE set, while the 
weak--hypercharge does not possess the canonical $SO(10)$ embedding. 
We remark that this type of enhanced symmetry also plays a roll
in forbidding operators which mediate proton decay. We also
note that \LRS models without the canonical
$SO(10)$ embedding of the weak--hypercharge have also been
recently discussed in the framework of Type I string constructions 
\cite{ibaneztypeI}. There it was argued that the non--canonical
embedding of the weak--hypercharge is advantageous for 
obtaining coupling unification in that framework. In the 
heterotic string the natural unification scale is of course 
the GUT or the heterotic string scale, and therefore the
natural embedding of the weak--hypercharge is the canonical
one. Nevertheless, as we stated above, the main aim of
our exercise here is to demonstrate how the Standard Model
spectrum may still arise from $SO(10)$ representations
while the normalization of the weak--hypercharge (and consequently
the Weinberg angle at the unification scale) do not have the
canonical $SO(10)$ value. The usefulness of this result to
string models with a lower unification scale will then depend
on improved understanding of the duality relation between 
the various models and the properties which are maintained
in an extrapolation from weak to strong coupling. That is,
one can imagine that a property of the type we describe
here will have its correspondence also in the dual Type I models. 

One option is to define the weak--hypercharge
with the standard $SO(10)$ embedding, as in eq.~(\ref{U1Y}),
\beq
          U(1)_Y~=~{1\over3}\,U(1)_C~+~{1\over2}\,U(1)_L~.
\label{U1Y}
\eeq
This is identical to the weak--hypercharge definition in 
$SU(3)\times SU(2)\times U(1)_Y$  free fermionic models, which
do not have enhanced symmetries.
However, in the present model, the $U(1)_C$ symmetry is now part
of the extended custodial symmetry $SU(2)_{\rm cust}$.  Expressing
$U(1)_C$ in terms of the new linear combinations defined above, we have
\beq
    {1\over3}\,U(1)_{C}~=~{1\over6}\biggl\lbrace 
{1\over4}( 3 T^3_{\rm cust}+U_{7^{\prime\prime}})-
{1\over7}( 3 U_{3^\prime}+U_{8^{\prime\prime}})\biggr\rbrace~.
\label{U1Cin274}
\eeq
Thus $U(1)_Y$, by depending on $T^3_{\rm cust}$, 
is no longer orthogonal to $SU(2)_{\rm cust}$.
We must therefore instead define the new linear combination with this term
removed,
\beqn
        U(1)_{Y'} &\equiv& U(1)_Y - {1\over {2}}\,T^3 \nonumber\\
        &=& {1\over 2}\,U(1)_L + {5\over{24}}\,U(1)_C \nonumber\\
         &&~~~~~-{1\over{8}}\,\biggl\lbrack
U(1)_1+U(1)_2+U(1)_3-U(1)_7-U(1)_9\biggr\rbrack~,
\label{U1pin274}
\eeqn
so that the weak--hypercharge is expressed in terms of $U(1)_{Y'}$ as
\beq
        U(1)_{Y} = U(1)_{Y'} + {1\over {2}}\,T^3_{\rm cust} 
~~~\Longrightarrow~~~~
           Q_{\rm e.m.} = T^3_L + Y = T^3_L + Y' + {1\over{2}}\,T^3_{\rm
cust}~.\label{Qemin274}
\eeq
The final observable gauge group then takes the form
\beq
       SU(3)_C \times SU(2)_L \times SU(2)_R \times SU(2)_{\rm cust}
\times U(1)_{Y'} ~\times
        ~\biggl\lbrace ~{\rm seven~other~}U(1){\rm ~factors}~\biggr\rbrace~.
\label{finalgroup}
\eeq
The remaining seven $U(1)$ factors must be chosen
as linear combinations of the previous $U(1)$ factors so as to be orthogonal
to the each of the other factors in (\ref{finalgroup}).

Next we discuss the Ka\v{c}--Moody factors associated with
the $U(1)$ factors in this model.
In this class of string models,
the Ka\v{c}--Moody level of the non--Abelian group factor is always one.
The situation is somewhat more complicated for the $U(1)$ factors, however.
In general, a given $U(1)$ current $U$ will be a combination of the simple
worldsheet
$U(1)$ currents $U_f\equiv f^*f$ corresponding to individual worldsheet
fermions $f$,
and will take the form $U=\sum_f a_f U_f$ where the $a_f$ are certain
model--specific coefficients.
The $U_f$ are each individually normalized to one, so that
$\langle U_f, U_f\rangle=1$. To produce the correct conformal dimension
for the massless states, each of the $U(1)$ linear combinations $U$ must also
be normalized to one.
The proper normalization coefficient for the linear combination
$U$ is thus given by $N=(\sum_f a_f^2)^{-1/2}$, so that
the properly normalized $U(1)$ current ${\hat U}$ is
given by ${\hat U}=N\cdot U$.

Now in general, the Ka\v{c}--Moody level of the $U(1)_Y$ generator
can be deduced from the OPE's between two of the $U(1)$ currents,
and will be
\beq
       k_1~=~ 2\,N^{-2}~ = ~ 2\,\sum_f \, a_f^2~.
\label{k1u1}
\eeq
For a weak--hypercharge that is a combination
of several $U(1)$'s with {\it different}\/ normalizations,
the result (\ref{k1u1}) generalizes to
\beq
       k_1 ~=~ \sum_i\, a_i^2 \, k_i~
\label{ky}
\eeq
where the $k_i$ are the individual normalizations
for each of the $U(1)$'s.

In Model 1, the $U(1)_Y$ generator is given as a
combination of simple worldsheet
currents that produces the correct weak--hypercharges for the Standard Model
particles. Thus, in that case, $k_1$ is simply given by (\ref{k1u1}).
However, for the weak--hypercharges (\ref{U1pin274}) and (\ref{Qemin274})
that appear in Model 2 we instead use (\ref{ky}).
Hence, for this weak--hypercharge, we see from (\ref{Qemin274}) and (\ref{ky})
that $k_1=(1/4)k_{2_C}+k_{Y^\prime}=1/4+17/12=5/3$, which is the same as the
standard $SO(10)$ normalization.

Alternatively, we can define the weak--hypercharge to be the combination
\beq
         U(1)_Y~=~{1\over2}\,U(1)_L~-~ {1\over6}U(1)_{3^\prime}~+~
{1\over6} U_{7^{\prime\prime}}
\eeq
where $U(1)_{3^\prime}$ and $U(1)_{7^{\prime\prime}}$ 
are given in (\ref{u1com}). This combination still reproduces the
correct charge assignment for the Standard Model states.
In this case the Ka\v{c}--Moody levels of
$U(1)_L$, $U(1)_{3^\prime}$ and $U(1)_{7^{\prime\prime}}$ 
are 4, 28 and 48 respectively, so that $k_Y=28/9$.
Therefore, the Weinberg angle at the unification scale is 
$\sin^2\theta_W=0.243$. Naturally, the point that we
want to raise is not that the present model with this value
of $\sin^2\theta_W$ provides a realistic unified model. Rather,
we make the following interesting observation: 
The three Standard
Model generations still arise from $SO(10)$ representations.
Specifically,  the Standard Model three generations all arise
{}from the three $\mbf 16$ representations of $SO(10)$ of the NAHE set
basis vectors. However, the weak--hypercharge does not possess 
the standard $SO(10)$ embedding and consequently, $\sin^2\theta_W\ne3/8$
at the unification scale. Of course, it will be of further interest
to see if such a structure can also emerge from Type I string constructions
which actually allow for a lower unification scale. The results
of ref.~\cite{ibaneztypeI}, which show that 
some of the structure of compactifying the heterotic string
on a particular orbifold, is actually preserved also in the
Type I models, give rise to the suspicion that this may indeed be the case.

\section{Anomalous $U(1)$}\label{anomalousu1}

A general property of the realistic free fermionic heterotic string models,
which is also shared by many other superstring vacua, is the
existence of an ``anomalous" $U(1)$.
The presence of an Abelian anomalous symmetry in superstring
derived models yields many desirable phenomenological consequences
{}from the point of view of the effective low energy field theory.
Indeed, the existence of such an anomalous $U(1)$ symmetry
in string derived models has inspired vigorous attempts to 
understand numerous issues, relevant for the observable phenomenology, 
including: the fermion mass spectrum, supersymmetry breaking
cosmological implications, and more. From the perspective
of string phenomenology an important function of the
anomalous $U(1)$ is to induce breaking and rank reduction
of the four dimensional gauge group. In general, the existence
of an anomalous $U(1)$ in a string model implies that the
string vacuum is unstable and must be shifted 
to a stable point in the moduli space.
This arises because, by the Green--Schwarz
anomaly cancellation mechanism, the anomalous $U(1)$ gives rise
to a Fayet--Iliopoulos term which breaks supersymmetry. 
Supersymmetry is restored and the vacuum is stabilized by sliding the
vacuum along flat $F$ and $D$ directions. This is achieved by assigning
non--vanishing VEVs to some scalar fields in the massless string spectrum.

An important issue in string phenomenology is therefore to
understand what are the general conditions for the appearance
of an anomalous $U(1)$ and under what conditions an anomalous
$U(1)$ is absent. The previously studied realistic free fermionic 
string models, which include the \FSUc, \PSc, and \SLM
types, have always contained an anomalous $U(1)$ symmetry.
In contrast, in the two \LRS models defined respectively by 
(\ref{model1},\ref{phasesmodel1})
and (\ref{model2},\ref{phasesmodel2}) all the $U(1)$ symmetries in the four
dimensional gauge group are anomaly free. This is, in fact, the first instance 
that realistic three generation (2,0) heterotic string models have produced 
models which do not contain an anomalous $U(1)$ symmetry. Irrespective of the
potential phenomenological merit of an anomalous $U(1)$ symmetry,
it is important to extract the properties of the models
that result in the presence, or the
absence, of an anomalous $U(1)$ symmetry.

For completeness we first discuss the case of the free fermionic
models which contain an anomalous $U(1)$, {\it i.e.},
the \FSUc, the \PSc, and the \SLM string models.
The question of the anomalous $U(1)$ symmetry in string models, in general,
and in the free fermionic models, in particular, was studied in some detail
in ref.~\cite{cf1}. The anomalous $U(1)$ in the free fermionic models
is in general a combination of two distinct kinds of world--sheet $U(1)$
currents,
those generated by ${\bar\eta}^j$ and those generated by the additional
complexified fermions from the set $\{{\bar y},{\bar\omega}\}^{1,\cdots,6}$.
The trace of the $U(1)$ charges of the entire massless string spectrum
can then be non--vanishing under some of these world--sheet $U(1)$ currents.
One combination of these $U(1)$ currents then becomes the anomalous
$U(1)$, whereas all the orthogonal combinations are anomaly free.
To understand the origin of the anomalous $U(1)$ in the realistic
free fermionic models, it is instructive to consider the contributions
{}from the two types of world--sheet $U(1)$ currents separately.

In ref.~\cite{cf1} it was shown that the anomalous $U(1)$ 
in the realistic free fermionic models can be seen to arise
due to the breaking of the world--sheet supersymmetry from
(2,2) to (2,0). Consider the set of boundary condition
basis vectors $\{{\bf1},\mS, \mzeta,\mX,\mb_1,\mb_2\}$ \cite{cf1},
which produces (for an appropriate choice of the GSO phases)
the model with $SO(12)\times E_6\times U(1)^2\times E_8$ gauge
group. It was shown that if we choose the GSO phases such
that $E_6\rightarrow SO(10)\times U(1)$, the $U(1)$ in the
decomposition of $E_6$ under $SO(10)\times U(1)$ becomes
the anomalous $U(1)$. This $U(1)$ is produced by the
combination of world--sheet currents ${\bar\eta}^1{\bar\eta}^{1^*}+
{\bar\eta}^2{\bar\eta}^{2^*}+{\bar\eta}^3{\bar\eta}^{3^*}$.
We can view all of the realistic \FSUc, \PSc, and \SLM free fermionic
string models as being related to this $SO(12)\times 
E_6\times U(1)^2\times E_8$ string vacuum. This combination of
$U(1)$ currents therefore contributes to the anomalous $U(1)$ 
in all the realistic free fermionic models with
\FSUc, \PSc, or \SLM gauge groups. 

The existence of the anomalous $U(1)$ in the \FSUc, \PSc, or \SLMc,
and its absence in the \LRS string models
can be traced to different $N=4$ string vacua in four
dimensions. While in the $E_6$ model one starts with  
an $N=4$ $SO(12)\times E_8\times E_8$ string vacua,
produced by the set $\{{\bf1},\mS,\mX,\mzeta\}$ \cite{cf1},
we can view the \FSUc, \PSc, and \SLM string models as
starting from an $N=4$ $SO(12)\times SO(16)\times SO(16)$ string vacua. 
In this case the two spinorial representations from the sectors
$\mX$ and $\mzeta$, that complete
the adjoint of $SO(16)\times SO(16)$ to $E_8\times E_8$, are
projected out by the choice of the GSO projection phases. 
The subsequent projections, induced by the basis vectors 
$\mb_1$ and $\mb_2$, which correspond to the $\IZ_2\times \IZ_2$
orbifold twistings, then operate identically in the two models, 
producing in one case the $E_6$, and in the second case 
the $SO(10)\times U(1)$, gauge groups, respectively.
The important point, however, is that both cases preserve
the ``standard embedding'' structure which splits the
observable and hidden sectors. The important set in this 
respect is the set $\{{\bf1},\mS,\mX,\mzeta\}$, where $\mX$
has periodic boundary conditions for $\{{\bar\psi}^{1,\cdots,5},
{\bar\eta}^1,{\bar\eta}^2,{\bar\eta}^3\}$. The choice of
the phase $c{\mX\choose\mzeta}=\pm1$ fixes the vacuum to 
$E_8\times E_8$ or $SO(16)\times SO(16)$.

In contrast,
the \LRS free fermionic string models do not start
with the $N=4$ $E_8\times E_8$ or $SO(16)\times SO(16)$ vacua. 
Rather, in this case the starting $N=4$ vacua can be seen to arise
{}from the set of boundary condition basis vectors
$\{{\bf1},\mS,2\mgamma,\mX\}$. 
Starting with this set and with the choice of GSO projection phases
\begin{equation}
{\bordermatrix{
              &{\bf 1}& \mS & & \mX    & 2\mgamma \cr
       {\bf 1}&~~1&~~1 & & -1   &  -1     \cr
           \mS&~~1&~~1 & & -1   &  -1     \cr
	      &   &    & &      &         \cr
          \mX & -1& -1 & &~~1   & ~~1     \cr
     2\mgamma & -1& -1 & & -1   & ~~1     \cr}}\, ,
\label{phasesn4model}
\end{equation}
the resulting string vacua has $N=4$ space--time supersymmetry
with $SO(16)\times E_7\times E_7$ gauge group. 
The sectors $\mb_1$ and $\mb_2$ are then added as
in the previous models. The \LRS string models therefore
do not preserve the ``standard embedding'' splitting between the
observable and hidden sectors. This is the first basic
difference between the \FSUc, \PSc, or \SLMc,
and the \LRS free fermionic models. 

Now turn to the case of the three generation models. The chirality
of the generations from the sectors $\mb_j$ $(j=1,2,3)$ is induced
by the projection which breaks $N=2\rightarrow N=1$ space--time
supersymmetry. Chirality for the generations is therefore fixed by the GSO
projection phase $c{\mb_i\choose{\mb_j}}$ with $i\ne j$. 
On the other hand,  
generation charges under $U(1)_j$ are fixed
by the $\mX$ projection in the $E_6$ model, by
the projection induced by the vector $2\mgamma$
of the \FSUc, \PSc, and \SLM string models, or by the vector
$2\mgamma$ of the \LRS string models. The difference
is that in the case of the \FSUc, \PSc, and \SLM string models 
the $2\mgamma$ projection fixes the same sign for the 
$U(1)_j$ charges of the states from the sectors $\mb_j$. 
In contrast, in the \LRS free fermionic models the corresponding 
$2\mgamma$ projection fixes one sign for the $(Q_R+L_R)_j$ 
states and the opposite sign for the $(Q_L+L_L)_j$ states. 
The consequence is that the total trace vanishes and the 
sectors $\mb_j$ do not contribute to the trace of the $U(1)_j$ charges.
This is in fact the reason that \LRS free fermionic 
models can appear without an anomalous $U(1)$.

We stress that the existence of \LRS free fermionic string models
without an anomalous $U(1)$ does not preclude the possibility of
other \LRS models with an anomalous $U(1)$. Our Model 3, specified by
Table (\ref{model3}) and matrix (\ref{phasesmodel3}) provides a 
counter--example.
\vskip 0.4truecm

LRS Model 3 Boundary Conditions:
\beqn
 &\begin{tabular}{c|c|ccc|c|ccc|c}
 ~ & $\psi^\mu$ & $\chi^{12}$ & $\chi^{34}$ & $\chi^{56}$ &
        $\bar{\psi}^{1,...,5} $ &
        $\bar{\eta}^1 $&
        $\bar{\eta}^2 $&
        $\bar{\eta}^3 $&
        $\bar{\phi}^{1,...,8} $ \\
\hline
\hline
 ${\malpha}$  &  0 & 0&0&0 & 1~1~1~0~0 & 0 & 0 & 0 &1~1~1~1~0~0~0~0 \\
 ${\mbeta}$   &  0 & 0&0&0 & 1~1~1~0~0 & 0 & 0 & 0 &1~1~1~1~0~0~0~0 \\
 ${\mgamma}$  &  0 & 0&0&0 &
${1\over2}$~${1\over2}$~${1\over2}$~0~0&${1\over2}$&${1\over2}$&${1\over2}$ &
\end{tabular}
   \nonumber\\
   ~  &  ~ \nonumber\\
   ~  &  ~ \nonumber\\
     &\begin{tabular}{c|c|c|c}
 ~&   $y^3{y}^6$
      $y^4{\bar y}^4$
      $y^5{\bar y}^5$
      ${\bar y}^3{\bar y}^6$
  &   $y^1{\omega}^5$
      $y^2{\bar y}^2$
      $\omega^6{\bar\omega}^6$
      ${\bar y}^1{\bar\omega}^5$
  &   $\omega^2{\omega}^4$
      $\omega^1{\bar\omega}^1$
      $\omega^3{\bar\omega}^3$
      ${\bar\omega}^2{\bar\omega}^4$ \\
\hline
\hline
$\malpha$& 1 ~~~ 1 ~~~ 1 ~~~ 0  & 1 ~~~ 1 ~~~ 1 ~~~ 0  & 1 ~~~ 1 ~~~ 1 ~~~ 0 \\
$\mbeta$ & 0 ~~~ 1 ~~~ 0 ~~~ 1  & 0 ~~~ 1 ~~~ 0 ~~~ 1  & 1 ~~~ 0 ~~~ 0 ~~~ 0 \\
$\mgamma$& 0 ~~~ 0 ~~~ 1 ~~~ 1  & 1 ~~~ 0 ~~~ 0 ~~~ 0  & 0 ~~~ 1 ~~~ 0 ~~~ 1 \\
\end{tabular}
\label{model3}
\eeqn

LRS Model 3 Generalized GSO Coefficients:
\begin{equation}
{\bordermatrix{
          &{\bf 1}&\mS & &{\mb_1}&{\mb_2}&{\mb_3}& &{\malpha}&{\mbeta}&{\mgamma}\cr
       {\bf 1}&~~1&~~1 & & -1   &  -1 & -1  & & ~~1     & ~~1   & ~~i   \cr
           \mS&~~1&~~1 & &~~1   & ~~1 &~~1  & &  -1     &  -1   &  -1   \cr
	      &   &    & &      &     &     & &         &       &       \cr
       {\mb_1}& -1& -1 & & -1   &  -1 & -1  & &  -1     &  -1   & ~~i   \cr
       {\mb_2}& -1& -1 & & -1   &  -1 & -1  & &  -1     &  -1   & ~~i   \cr
       {\mb_3}& -1& -1 & & -1   &  -1 & -1  & &  -1     & ~~1   & ~~i   \cr
	      &   &    & &      &     &     & &         &       &       \cr
     {\malpha}&~~1& -1 & &~~1   & ~~1 &~~1  & & ~~1     & ~~1   & ~~1   \cr
      {\mbeta}&~~1& -1 & & -1   &  -1 &~~1  & &  -1     &  -1   &  -1   \cr
     {\mgamma}&~~1& -1 & &~~1   &  -1 &~~1  & &  -1     &  -1   & ~~1   \cr}}
\label{phasesmodel3}
\end{equation}

Similar to our Models 1 and 2, Model 3
uses the \LRS breaking pattern. 
It also contains three generations
{}from the sectors $\mb_1$, $\mb_2$ and $\mb_3$, and the untwisted 
spectrum is similar to that of the previous two models. 
However, Model 3 actually contains three anomalous 
$U(1)$ symmetries: Tr${U_4}=-24$, Tr${U_5}=-24$, Tr${U_6}=-24$,
one combination of which,
\beq
U(1)_A=U_4+U_5+U_6\label{u1a},
\eeq
is anomalous, while two orthogonal combinations
are anomaly free. In this model we see that the anomalous
$U(1)$'s correspond to $U(1)$ symmetries which
arise from the additional complexified world--sheet fermions
in the set $\{{\bar y},{\bar\omega}\}^{1,\cdots,6}$. 
This is in agreement with the above argument that the
$U(1)_{j=1,2,3}$, which are generated by the ${\bar\eta}^j$
world--sheet fermions, are anomaly free in the \LRS
free fermionic string models. The potential implications
for the flavor mass spectrum are of particular interest
in this regard. 

To close the discussion on the anomalous $U(1)$ in the the \LRS
string models we remark that in ref.~\cite{cf1} the general 
conditions that forbid the appearance of anomalous $U(1)$
in string models were discussed.  
Theorem 3, part $(b)$, of \cite{cf1} allows us to prove, 
without computing the trace of the charge
${\rm Tr}\, Q^{i}$, for each $U(1)_i$,   
that {\it all} of the Abelian gauge groups in
a given model are anomaly free. Theorem 3 states that:

{\narrower\smallskip\noindent  
A model is completely free of anomalous $U(1)$ if, for each $U(1)_i$, there
is at least one simple gauge group $\cal G$ for which
(a) all non--trivial massless reps of $\cal G$ do not carry $U(1)_i$ charge,
or
(b) the trace of $Q_i$ over all massless non--trivial reps of $\cal G$ 
is zero.\smallskip}

In Model 1 the only two non--trivial representations of the hidden sector
$SU(3)_{H_1}$ gauge group is the vector--like pair of fields,
$H_1$ and $\overline{H}_1$. 
Thus, the trace of each $U(1)_i$ charge over
$SU(3)_{H_1}$ reps is clearly zero. This implies, by Theorem 3(b), 
that all of the $U(1)_i$ in this model are anomaly free.  
(Note that the vector--like fields $H_2$ and $\overline{H}_2$ of $SU(3)_{H_2}$
imply this also. Furthermore,
the same method of proof also applies to Model 2 
when we replace $SU(3)_{H_1,H_2}$ with $SU(4)_{H_1,H_2}$.)

{\oddsidemargin  10.5pt \evensidemargin  10.5pt
\textheight  612pt \textwidth  432pt
\headheight  12pt \headsep  20pt
\footheight  12pt \footskip  40pt
\section{Phenomenological discussion and conclusions}\label{conclusion}
Clearly the most striking new feature of the \LRS
 string models that we presented is the absence
of an anomalous $U(1)$ symmetry in two of them. 
In the past much of the 
analysis of the three generation free fermionic models 
involved the analysis of flat directions that are induced
by the cancelation of the anomalous $U(1)$ D--term
\cite{flipped,ln1}.  
The Standard Model singlet VEVs that are used to cancel this
D--term spontaneously break some of the additional $U(1)$ symmetries
in the string models. Such singlet
VEVs are necessitated by the requirement that the superstring
vacuum preserves supersymmetry at the string scale. 
An important issue in this regard is therefore whether for such flat
directions there exist one, or perhaps several, $U(1)$ 
combinations which remain unbroken. 
$U(1)$ combinations that remain unbroken down to sufficiently
low energies may give rise to interesting observable effects.
In contrast, the absence of an anomalous $U(1)$ in the
LRS string models that we discussed here
allows, in principle, all the extra $U(1)$'s to remain 
unbroken at the string scale. Of course imposing some
phenomenological constraints, like the decoupling of 
fractionally charged states, may force some Standard Model
singlet fields to acquire a non--vanishing VEV. In which 
case the analysis of the flat directions is reintroduced. 
However, this is not necessitated by the requirement that
the anomalous $U(1)$ D--term vanishes, and hence not by
the requirement that the vacuum preserves supersymmetry.
Similarly, the Fayet--Iliopoulos term which is induced
by the anomalous $U(1)$ \cite{dsw} gives rise to an order parameter
which, together with the flat direction singlet VEVs,
is used to produce the hierarchical fermion mass pattern.
This order parameter is therefore no longer present if
there is no anomalous $U(1)$, as is the case in some 
of the \LRS string models that we
presented.

Examining the field content of Model 1, we note that the model
contains the neutral Standard Model singlet component of the
$\mbf 16$ of $SO(10)$. This field can therefore be used to break the
$SU(2)_R$ gauge symmetry. However, it is noted that the corresponding 
component of the $\overline{\mbf 16}$ is absent from the model
and all the remaining $SU(2)_R$ doublets have fractional
electric charge with respect to the most natural definition
of the weak--hypercharge give in eq.~(\ref{U1Y}).
The absence of the corresponding neutral component from
the $\overline{\mbf 16}$ is a common feature in the three generation
free fermionic models in which the $SO(1O)$ is broken
by at least two different basis vectors. 
Therefore, assuming that supersymmetry is not broken
at a high scale, the $SU(2)_R$ symmetry cannot be broken along a
supersymmetric flat directions, and can only be broken
by a VEV that does not preserve supersymmetry at a lower
scale. From Table 2 we note that
Model 2 contains two $SU(2)_C$ doublets, $N_{\alpha\beta}$
and $\overline{N}_{\alpha\beta}$, from the sector
$\mS+\mb_1+\mb_2+\malpha+\mbeta$, that can be used to break the
custodial $SU(2)$ symmetry along a flat direction. 

As with Model 1, both Models 2 and 3 contain the neutral 
component of the $\mbf 16$ of $SO(10)$ that can be employed
to break the $SU(2)_R$ symmetry.
We remark that this conclusion holds for the definition of 
the weak--hypercharge as given in eq.~(\ref{U1Y}). Other
viable definitions of the weak--hypercharge may result
in more electrically neutral fields that may be used
to break $SU(2)_R$. As we discussed above, with such
possible alternative definitions the Standard Model
spectrum still arises from the ${\mbf 16}$ representation of $SO(1O)$,
but the weak--hypercharge normalization differs from the
canonical $SO(10)$ normalization. 

{}From Tables 1--3 we note that all three models contain the
required Higgs bi--doublet representations, $h_1$ and $h_2$,
that are needed in order to generate the Standard Model gauge
boson and fermion masses. Examining the superpotential terms,
eqs.~(\ref{w3ssm2a}) and (\ref{w3ssm}), 
of Models 1 and 2 respectively,  we note that the couplings
$Q_{L_i}Q_{R_i}h_i$ and $L_{L_i}L_{R_i}h_i$ exist to provide
potential mass terms for the states from the sectors ${\mb}_1$ 
and ${\mb}_2$. The structure of the basis vectors beyond the
NAHE set, $\malpha$, $\mbeta$ and $\mgamma$, which break the cyclic
permutation symmetry between the three twisted sectors $\mb_1$, 
$\mb_2$ and $\mb_3$, results in the states from the sector
$\mb_3$ being identified with the lightest generation.
This outcome is similar to the result that was found in the
case of the free fermionic \SLM string models \cite{lfm}, 
and again is a reflection of the breaking of the cyclic 
permutation symmetry by the basis vectors $\malpha$, $\mbeta$ and
$\mgamma$.

>From eq.~(\ref{w3hid2a}) we note that in Model 1, provided that
$\Phi_3$ gets a non--vanishing VEV of the order of the string scale,
then the entire hidden matter spectrum of Model 1 becomes superheavy. 
In this case the content of the hidden sector spectrum consists
of the gauge bosons of the unbroken hidden $E_8$ subgroup,
which in model 1 is $SU(3)\times SU(3)\times U(1)^4$. 
These hidden states can interact with the observable 
sector only via the heavy hidden matter states,
which are charged with respect to the horizontal 
$U(1)$ symmetries, $U(1)_{1,2}$. The observable sector
states are also charged under the horizontal $U(1)_{1,2}$ 
symmetries. This represents the interesting case that the 
lightest hidden sector state is a hidden glueball that can
interact with the Standard Model states only via the superheavy
fermions. Such states may provide interesting
dark matter candidates \cite{fp}. 

In this paper we extended the case studies of realistic
free fermionic string models to the case in which the observable
universal $SO(10)$ gauge group is broken to the left--right symmetric, 
$SU(3)_C\times SU(2)_L\times SU(2)_R\times U(1)_{B-L}$, 
gauge group. We presented three specific examples with this 
symmetry breaking pattern together with the entire superpotential
terms for the first two models, up to quintic order. 
The distinctive feature of the LRS free fermionic string models,
as compared to the previous, FSU5, PS, and SLM cases, 
is the existence of models which do not contain an anomalous
$U(1)$ symmetry. We discussed the general structures which
result in the absence, or presence, of an anomalous $U(1)$,
in the respective cases. 
We further contemplated how the string models can motivate the interesting
possibility in which the Standard Model fermion spectrum
arises from three $\mbf 16$ representations of $SO(10)$,
while the weak--hypercharge does not possess the canonical
$SO(10)$ embedding. Finally, it will be of further interest
to study compactification of other classes of string theories
on the manifolds which are associated with the free fermionic
models and to examine the properties of the models, which
are preserved in these dual constructions. Similarly, it is
of further interest to study the properties of the LRS string models
in relation to the phenomenological studies of LRS field theory
models. We shall return to these and related questions in 
future publications. 

\newpage
\section{Acknowledgments}

This work is supported in part by DOE Grant No. DE--FG--02--94ER40823
(AEF,CS) and a PPARC advanced fellowship (AEF); and by DOE Grant
No. DE--FG--0395ER40917 (GBC).}

\appendix
{\textwidth=7.5in
\oddsidemargin=-18mm
\topmargin=-5mm
\renewcommand{\baselinestretch}{1.3}

{\textwidth=7.5in
\oddsidemargin=-18mm
\topmargin=-5mm
\renewcommand{\baselinestretch}{1.3}
\smallskip

\begin{flushleft}
\begin{table}
{\rm \large\bf A Left--Right~Symmetric~Model~1~Fields}
\begin{eqnarray*}
\begin{tabular}{|c|c|c|rrrrrrr|c|rrrr|}
\hline
  $F$ & SEC & $(C;L;R)$ & $Q_C$
   & $Q_1$ & $Q_2$ & $Q_3$
   & $Q_4$ & $Q_5$ & $Q_6$
   & $SU(3)_{H_{1;2}}$
   & $Q_7$ & $Q_8$ & $Q_9$ & $Q_{10}$ \\
\hline

   $Q_{L_1}$ & $\mb_1$ & $(3,2,1)$ & 2 &
       -2 &    0 &    0 &   -2 &    0 &    0 &
   $(1,1)$&    0 &    0 &    0 &    0 \\

   $Q_{R_1}$ &       & $({\bar3},1,2)$ & -2 &
        2 &    0 &    0 &    2 &    0 &    0 &
   $(1,1)$&    0 &    0 &    0 &    0 \\

   $L_{L_1}$ &       & $(1,2,1)$ & -6 &
       -2 &    0 &    0 &    2 &    0 &    0 &
   $(1,1)$&    0 &    0 &    0 &    0 \\

   $L_{R_1}$ &       & $(1,1,2)$ & 6 &
        2 &    0 &    0 &   -2 &    0 &    0 &
   $(1,1)$&    0 &    0 &    0 &    0 \\
\hline

   $Q_{L_2}$ & $\mb_2$ & $(3,2,1)$ & 2 &
        0 &   -2 &    0 &    0 &   -2 &    0 &
   $(1,1)$&    0 &    0 &    0 &    0 \\

   $Q_{R_2}$ &       & $({\bar3},1,2)$ & -2 &
        0 &    2 &    0 &    0 &    2 &    0 &
   $(1,1)$&    0 &    0 &    0 &    0 \\

   $L_{L_2}$ &       & $(1,2,1)$ & -6 &
        0 &   -2 &    0 &    0 &    2 &    0 &
   $(1,1)$&    0 &    0 &    0 &    0 \\

   $L_{R_2}$ &       & $(1,1,2)$ & 6 &
        0 &    2 &    0 &    0 &   -2 &    0 &
   $(1,1)$&    0 &    0 &    0 &    0 \\
\hline

   $Q_{L_3}$ & $\mb_3$ & $(3,2,1)$ & 2 &
        0 &    0 &   -2 &    0 &    0 &    2 &
   $(1,1)$&    0 &    0 &    0 &    0 \\

   $Q_{R_3}$ &       & $({\bar3},1,2)$ & -2 &
        0 &    0 &    2 &    0 &    0 &   -2 &
   $(1,1)$&    0 &    0 &    0 &    0 \\

   $L_{L_3}$ &       & $(1,2,1)$ & -6 &
        0 &    0 &   -2 &    0 &    0 &   -2 &
   $(1,1)$&    0 &    0 &    0 &    0 \\

   $L_{R_3}$ &       & $(1,1,2)$ & 6 &
        0 &    0 &    2 &    0 &    0 &    2 &
   $(1,1)$&    0 &    0 &    0 &    0 \\
\hline
   $\cL{+}{L_1}$ & $\mb_1+\mbeta$ & $(1,2,1)$ & 3 &
       -1 &    1 &    1 &    0 &    2 &   -2 &
   $(1,1)$&    0 &   -6 &   -6 &    0 \\

   $\cL{+}{R_1}$ &  $\pm\mgamma$ & $(1,1,2)$ & 3 &
       -1 &    1 &    1 &    0 &   -2 &    2 &
   $(1,1)$&    0 &   -6 &   -6 &    0 \\

   $\cL{-}{L_1}$ &       & $(1,2,1)$ & -3 &
        1 &   -1 &   -1 &    0 &    2 &   -2 &
   $(1,1)$&    0 &    6 &    6 &    0 \\

   $\cL{-}{R_1}$ &       & $(1,1,2)$ & -3 &
        1 &   -1 &   -1 &    0 &   -2 &    2 &
   $(1,1)$&    0 &    6 &    6 &    0 \\
\hline
   $\cL{+}{L_2}$ & $\mb_2+\mbeta$ & $(1,2,1)$ & 3 &
        1 &   -1 &    1 &    2 &    0 &   -2 &
   $(1,1)$&    0 &   -6 &   -6 &    0 \\

   $\cL{+}{R_2}$ &  $\pm\mgamma$ & $(1,1,2)$ & 3 &
        1 &   -1 &    1 &   -2 &    0 &    2 &
   $(1,1)$&    0 &   -6 &   -6 &   0 \\

   $\cL{-}{L_2}$ &       & $(1,2,1)$ & -3 &
       -1 &    1 &   -1 &    2 &    0 &   -2 &
   $(1,1)$&    0 &    6 &    6 &    0 \\

   $\cL{-}{R_2}$ &       & $(1,1,2)$ & -3 &
       -1 &    1 &   -1 &   -2 &    0 &    2 &
   $(1,1)$&    0 &    6 &    6 &    0 \\
\hline
   $\cL{+}{L_3}$ & $\mb_3+\mbeta$ & $(1,2,1)$ & 3 &
        1 &    1 &   -1 &    2 &    2 &    0 &
   $(1,1)$&    0 &   -6 &   -6 &    0 \\

   $\cL{+}{R_3}$ &  $\pm\mgamma$ & $(1,1,2)$ & 3 &
        1 &    1 &   -1 &   -2 &   -2 &    0 &
   $(1,1)$&    0 &   -6 &   -6 &    0 \\

   $\cL{-}{L_3}$ &       & $(1,2,1)$ & -3 &
       -1 &   -1 &    1 &    2 &    2 &    0 &
   $(1,1)$&    0 &    6 &    6 &    0 \\

   $\cL{-}{R_3}$ &       & $(1,1,2)$ & -3 &
       -1 &   -1 &    1 &   -2 &   -2 &    0 &
   $(1,1)$&    0 &    6 &    6 &    0 \\
\hline
$\cL{+}{L_{12}}$ & ${\bf 1}+\mb_1+$  & $(1,2,1)$ & 0 &
        2 &    2 &    0 &    0 &    0 &    2 &
   $(1,1)$&    4 &    0 &    0 &   12 \\

$\cL{+}{R_{12}}$ & $\mb_2+2\mgamma$ & $(1,1,2)$ & 0 &
       -2 &   -2 &    0 &    0 &    0 &   -2 &
   $(1,1)$&    4 &    0 &    0 &   12 \\

$\cL{-}{L_{12}}$ & & $(1,2,1)$ & 0 &
        2 &    2 &    0 &    0 &    0 &    2 &
   $(1,1)$&   -4 &    0 &    0 &  -12 \\

$\cL{-}{R_{12}}$ &     & $(1,1,2)$ & 0 & 
       -2 &   -2 &    0 &    0 &    0 &   -2 &
   $(1,1)$&   -4 &    0 &    0 &  -12 \\

\hline

$\cL{+}{L_{13}}$ & ${\bf 1}+\mb_1+$     & $(1,2,1)$ & 0 &
        2 &    0 &    2 &    0 &   -2 &    0 &
   $(1,1)$&    4 &    0 &    0 &   12 \\

$\cL{+}{R_{13}}$ & $\mb_3+2\mgamma$     & $(1,1,2)$ & 0 &
       -2 &    0 &   -2 &    0 &    2 &    0 &
   $(1,1)$&    4 &    0 &    0 &   12 \\

$\cL{-}{L_{13}}$ & & $(1,2,1)$ & 0 &
        2 &    0 &    2 &    0 &   -2 &    0 &
   $(1,1)$&   -4 &    0 &    0 &  -12  \\

$\cL{-}{R_{13}}$ &     & $(1,1,2)$ & 0 &
       -2 &    0 &   -2 &    0 &    2 &    0 &
   $(1,1)$&   -4 &    0 &    0 &  -12 \\

\hline
\end{tabular}
\label{model1matter1}
\end{eqnarray*}\\
{Table 1: Model 1 fields. Charges have been multiplied by 4.}
\end{table}
\end{flushleft}
\hfill\vfill

\begin{flushleft}
\begin{table}
\begin{eqnarray*}
\begin{tabular}{|c|c|c|rrrrrrr|c|rrrr|}
\hline

  $F$ & SEC & $(C;L;R)$ & $Q_C$
   & $Q_1$ & $Q_2$ & $Q_3$
   & $Q_4$ & $Q_5$ & $Q_6$
   & $SU(3)_{H_{1;2}}$
   & $Q_7$ & $Q_8$ & $Q_9$ & $Q_{10}$ \\

\hline
$\cL{+}{L_{23}}$ & ${\bf 1}+\mb_2+$ & $(1,2,1)$ & 0 &
        0 &    2 &    2 &   -2 &    0 &    0 &
   $(1,1)$&    4 &    0 &    0 &   12 \\

$\cL{+}{R_{23}}$ & $\mb_3+2\mgamma$ & $(1,1,2)$ & 0 &
        0 &   -2 &   -2 &    2 &    0 &    0 &
   $(1,1)$&    4 &    0 &    0 &   12 \\

$\cL{-}{L_{23}}$ &     & $(1,1,2)$ & 0 &
        0 &    2 &    2 &   -2 &    0 &    0 &
   $(1,1)$&   -4 &    0 &    0 &  -12 \\

$\cL{-}{R_{23}}$ & & $(1,2,1)$ & 0 &
        0 &   -2 &   -2 &    2 &    0 &    0 &
   $(1,1)$&   -4 &    0 &    0 &  -12  \\

\hline
   $D_{\malpha\mbeta}$ & $\mS+\mbeta$ & $(3,1,1)$ & -1 &
       -1 &   -1 &   -1 &    2 &    2 &    2 &
   $(1,1)$&    0 &    6 &    6 &    0 \\

   ${\bar D}_{\malpha\mbeta}$ & $\pm\mgamma$ & $({\bar 3},1,1)$ & 1 &
        1 &    1 &    1 &   -2 &   -2 &   -2 &
   $(1,1)$&    0 &   -6 &   -6 &    0 \\
\hline

   $h_1$    & Neveu--  & $(1,2,2)$ & 0 &
        0 &    0 &   0 &    0 &    0 &    0 &
   $(1,1)$&    0 &   0 &    0 &    0 \\

   $h_2$    & Schwarz  & $(1,2,2)$ & 0 &
        0 &    0 &   0 &    0 &    0 &    0 &
   $(1,1)$&    0 &   0 &    0 &    0 \\

   $\Phi_1$ &          & $(1,1,1)$ & 0 &
        0 &    0 &    0 &    0 &    0 &   0 &
   $(1,1)$&    0 &    0 &    0 &    0 \\

   $\Phi_2$ &          & $(1,1,1)$ & 0 &
        0 &    0 &    0 &    0 &    0 &   0 &
   $(1,1)$&    0 &    0 &    0 &    0 \\

   $\Phi_3$ &          & $(1,1,1)$ & 0 &
        0 &    0 &    0 &    0 &    0 &   0 &
   $(1,1)$&    0 &    0 &    0 &    0 \\

   $\Phi_{12}$ &       & $(1,1,1)$ & 0 &
       -4 &    4 &   0 &    0 &    0 &    0 &
   $(1,1)$&    0 &    0 &    0 &    0 \\

   ${\bar\Phi}_{12}$ &       & $(1,1,1)$ & 0 &
        4 &   -4 &   0 &    0 &    0 &    0 &
   $(1,1)$&    0 &    0 &    0 &    0 \\

   $\Phi_{23}$ &       & $(1,1,1)$ & 0 &
        0 &   -4 &   4 &    0 &    0 &    0 &
   $(1,1)$&    0 &    0 &    0 &    0 \\

   ${\bar\Phi}_{23}$ &       & $(1,1,1)$ & 0 &
        0 &    4 &  -4 &    0 &    0 &    0 &
   $(1,1)$&    0 &    0 &    0 &    0 \\

   $\Phi_{31}$ &       & $(1,1,1)$ & 0 &
       -4 &    0 &   4 &    0 &    0 &    0 &
   $(1,1)$&    0 &    0 &    0 &    0 \\

   ${\bar\Phi}_{31}$ &       & $(1,1,1)$ & 0 &
        4 &    0 &  -4 &    0 &    0 &    0 &
   $(1,1)$&    0 &    0 &    0 &    0 \\

\hline
   $\x{1 }$& ${\bf 1}+\mS+\mb_3+$& $(1,1,1)$ 
     &  0 &    2 &   -2 &    0 &    0 &    0 &    0 &   
   $(1,1)$&    4 &   12 &  -12 &    4\\
 $\bx{1 }$& $\malpha+\mbeta$    & $(1,1,1)$ 
     &  0 &   -2 &    2 &    0 &    0 &    0 &    0 &  
   $(1,1)$&   -4 &  -12 &   12 &   -4\\
  $\x{2 }$&    & $(1,1,1)$ 
     &  0 &    2 &   -2 &    0 &    0 &    0 &    0 &   
   $(1,1)$&    4 &   12 &   12 &   -4\\
 $\bx{2 }$&    & $(1,1,1)$ 
     &  0 &   -2 &    2 &    0 &    0 &    0 &    0 &  
   $(1,1)$&   -4 &  -12 &  -12 &    4\\
  $\x{3 }$&    & $(1,1,1)$ 
     &  0 &    2 &   -2 &    0 &    0 &    0 &    0 &  
   $(1,1)$&   -4 &  -12 &  -12 &    4\\
 $\bx{3 }$&    & $(1,1,1)$ 
     &  0 &   -2 &    2 &    0 &    0 &    0 &    0 &  
   $(1,1)$&    4 &   12 &   12 &   -4\\
  $\x{4 }$&    & $(1,1,1)$ 
     &   0 &    2 &  -2 &    0 &    0 &    0 &    0 &
   $(1,1)$&   -4 &  -12 &   12 &   -4\\
 $\bx{4 }$&    & $(1,1,1)$ 
     &   0 &   -2 &   2 &    0 &    0 &    0 &    0 &
   $(1,1)$&    4 &   12 &  -12 &    4\\
\hline
   $\x{5 }$& $\mS+\mbeta$     & $(1,1,1)$ & -3 &
       -3 &    1 &    1 &    2 &    2 &   -2 &
   $(1,1)$&    0 &   -6 &   -6 &    0 \\

   $\bx{5 }$& $\pm\mgamma$     & $(1,1,1)$ & 3 &
        3 &   -1 &   -1 &   -2 &   -2 &    2 &
   $(1,1)$&    0 &    6 &    6 &    0 \\

   $\x{6 }$&     & $(1,1,1)$ & -3 &
        1 &   -3 &    1 &    2 &    2 &   -2 &
   $(1,1)$&    0 &   -6 &   -6 &    0 \\

   $\bx{6 }$&     & $(1,1,1)$ & 3 &
       -1 &    3 &   -1 &   -2 &   -2 &    2 &
   $(1,1)$&    0 &    6 &    6 &    0 \\

   $\x{7 }$&     & $(1,1,1)$ & -3 &
        1 &    1 &   -3 &    2 &    2 &   -2 &
   $(1,1)$&    0 &   -6 &   -6 &    0 \\

   $\bx{7 }$&     & $(1,1,1)$ & 3 &
       -1 &   -1 &    3 &   -2 &   -2 &    2 &
   $(1,1)$&    0 &    6 &    6 &    0 \\

\hline
\end{tabular}
\nolabel
\end{eqnarray*}
{Table 1 continued: Model 1 fields.}
\end{table}
\end{flushleft}

\begin{flushleft}
\begin{table}
\begin{eqnarray*}
\begin{tabular}{|c|c|c|rrrrrrr|c|rrrr|}
\hline
  $F$ & SEC & $(C;L;R)$ & $Q_C$
   & $Q_1$ & $Q_2$ & $Q_3$
   & $Q_4$ & $Q_5$ & $Q_6$
   & $SU(3)_{H_{1;2}}$
   & $Q_7$ & $Q_8$ & $Q_9$ & $Q_{10}$ \\
\hline

   $\x{8 }$  & ${\bf 1}+\mb_1+$ & $(1,1,1)$ & -3 &
        1 &    1 &    1 &   -2 &    2 &   -2 &
   $(1,1)$&    4 &    6 &    6 &   12 \\

   $\bx{8 }$ & $\mb_2+\mb_3+$ & $(1,1,1)$ & 3 &
       -1 &   -1 &   -1 &    2 &   -2 &    2 &
   $(1,1)$&   -4 &   -6 &   -6 &  -12 \\

   $\x{9 }$  & $\mbeta\pm\mgamma$& $(1,1,1)$ & -3 &
        1 &    1 &    1 &    2 &   -2 &   -2 &
   $(1,1)$&   -4 &    6 &    6 &  -12 \\

   $\bx{9 }$ &     & $(1,1,1)$ & 3 &
       -1 &   -1 &   -1 &   -2 &    2 &    2 &
   $(1,1)$&    4 &   -6 &   -6  &   12 \\
\hline
$\x{10}$ & $\mS+\mb_1+$ & $(1,1,1)$ & 6 &
        0 &    0 &   -2 &    0 &    0 &    0 &
   $(1,1)$&    0 &    4 &  -12 &    0 \\

$\bx{10}$ & $\mb_2+\malpha+$ & $(1,1,1)$ & -6 &
        0 &    0 &    2 &    0 &    0 &    0 &
   $(1,1)$&    0 &   -4 &   12 &    0 \\

$\x{11}$ & $\mbeta+2\mgamma$ & $(1,1,1)$ & 6 &
        0 &    0 &   -2 &    0 &    0 &    0 &
   $(1,1)$&    0 &   -4 &  -12 &    0 \\

$\bx{11}$ &     & $(1,1,1)$ & -6 &
        0 &    0 &    2 &    0 &    0 &    0 &
   $(1,1)$&    0 &    4 &   12 &    0 \\
\hline

   $H_1$ & $\mS+\mb_1$ & $(1,1,1)$ & 0 &
       -2 &   -2 &    0 &    0 &    0 &    0 &
   $(3,1)$& -4 &   -4 &   -4 &    4 \\

   ${\bar H}_1$ & $+\mb_2+\malpha$ & $(1,1,1)$ & 0 &
        2 &    2 &    0 &    0 &    0 &    0 &
  $(\bt,1)$& 4 &    4 &    4 &   -4 \\

   $H_2$ & $+\mbeta$ & $(1,1,1)$ & 0 &
        2 &    2 &    0 &    0 &    0 &    0 &
   $(1,3)$& -4  &   4 &    4 &    4 \\

   ${\bar H}_2$ & $\oplus \mzeta $   & $(1,1,1)$ & 0 &
       -2 &   -2 &    0 &    0 &    0 &    0 &
 $(1,\bt)$&  4 &   -4 &   -4 &   -4 \\

\hline
\end{tabular}
\nolabel
\end{eqnarray*}
{Table 1 continued: Model 1 fields.} 
\end{table}
\end{flushleft}
\vskip 8truein
\hfill\vfill\eject

}




\setcounter{section}{1}

{\textwidth=7.5in
\oddsidemargin=-18mm
\topmargin=-5mm
\renewcommand{\baselinestretch}{1.3}

\section{Left--Right Symmetric Model 1 Superpotential Terms} 
\no \underline{Model 1 Fifth Order Superpotential:}
\begin{flushleft}
\no $W_5({\rm observable})$:
\beqn 
{
\begin{tabular}{llll} 
   $\php \h{1 }  \Q{L_2}  \Q{R_2}  \x{1 } \bx{1 } $  
 & $+ \h{1 }  \Q{L_2}  \Q{R_2}  \x{2 } \bx{2 } $ 
 & $+ \h{1 }  \Q{L_2}  \Q{R_2}  \x{3 } \bx{3 } $  
 & $+ \h{1 }  \Q{L_2}  \Q{R_2}  \x{4 } \bx{4 } $ \\ 
   $+ \h{1 }  \Q{L_2}  \Q{R_2}  \x{10} \bx{10} $  
 & $+ \h{1 }  \Q{L_2}  \Q{R_2}  \x{11} \bx{11} $ 
 & $+ \h{1 }  \L{L_2}  \L{R_2}  \x{1 } \bx{1 } $  
 & $+ \h{1 }  \L{L_2}  \L{R_2}  \x{2 } \bx{2 } $ \\
   $+ \h{1 }  \L{L_2}  \L{R_2}  \x{3 } \bx{3 } $  
 & $+ \h{1 }  \L{L_2}  \L{R_2}  \x{4 } \bx{4 } $  
 & $+ \h{1 }  \L{L_2}  \L{R_2}  \x{10} \bx{10} $  
 & $+ \h{1 }  \L{L_2}  \L{R_2}  \x{11} \bx{11} $ \\
   $+ \h{1 } \cL{+}{L_{13}} \cL{-}{R_{13}}  \x{1 } \bx{1 } $  
 & $+ \h{1 } \cL{+}{L_{13}} \cL{-}{R_{13}}  \x{2 } \bx{2 } $  
 & $+ \h{1 } \cL{+}{L_{13}} \cL{-}{R_{13}}  \x{3 } \bx{3 } $  
 & $+ \h{1 } \cL{+}{L_{13}} \cL{-}{R_{13}}  \x{4 } \bx{4 } $ \\
   $+ \h{1 } \cL{+}{L_{13}} \cL{-}{R_{13}}  \x{10} \bx{10} $  
 & $+ \h{1 } \cL{+}{L_{13}} \cL{-}{R_{13}}  \x{11} \bx{11} $  
 & $+ \h{1 } \cL{-}{L_{13}} \cL{+}{R_{13}}  \x{1 } \bx{1 } $  
 & $+ \h{1 } \cL{-}{L_{13}} \cL{+}{R_{13}}  \x{2 } \bx{2 } $ \\
   $+ \h{1 } \cL{-}{L_{13}} \cL{+}{R_{13}}  \x{3 } \bx{3 } $  
 & $+ \h{1 } \cL{-}{L_{13}} \cL{+}{R_{13}}  \x{4 } \bx{4 } $  
 & $+ \h{1 } \cL{-}{L_{13}} \cL{+}{R_{13}}  \x{10} \bx{10} $  
 & $+ \h{1 } \cL{-}{L_{13}} \cL{+}{R_{13}}  \x{11} \bx{11} $ \\
   $+ \h{1 } \cL{+}{L_2} \cL{-}{R_2}  \x{1 } \bx{1 } $  
 & $+ \h{1 } \cL{+}{L_2} \cL{-}{R_2}  \x{2 } \bx{2 } $  
 & $+ \h{1 } \cL{+}{L_2} \cL{-}{R_2}  \x{3 } \bx{3 } $  
 & $+ \h{1 } \cL{+}{L_2} \cL{-}{R_2}  \x{4 } \bx{4 } $ \\
   $+ \h{1 } \cL{+}{L_2} \cL{-}{R_2}  \x{10} \bx{10} $  
 & $+ \h{1 } \cL{+}{L_2} \cL{-}{R_2}  \x{11} \bx{11} $  
 & $+ \h{1 } \cL{-}{L_2} \cL{+}{R_2}  \x{1 } \bx{1 } $  
 & $+ \h{1 } \cL{-}{L_2} \cL{+}{R_2}  \x{2 } \bx{2 } $ \\
\end{tabular}} 
\nolabel
\eeqn

\no $W_5({\rm observable})$ continued:
\beqn 
{\begin{tabular}{llll} 
   $+\cL{+}{L_{12}} \cL{+}{L_1} \cL{-}{R_{23}} \cL{-}{R_3}  \P{2 } $  
 & $+\cL{+}{L_{12}} \cL{-}{L_1} \cL{-}{R_{23}} \cL{+}{R_3}  \P{13} $  
 & $+\cL{-}{L_{12}} \cL{+}{L_{13}} \cL{-}{R_{12}} \cL{+}{R_{13}}  \P{1 } $  
 & $+\cL{-}{L_{12}} \cL{+}{L_{23}} \cL{-}{R_{12}} \cL{+}{R_{23}}  \P{2 } $ \\
   $+\cL{-}{L_{12}} \cL{+}{L_2} \cL{+}{R_{13}} \cL{-}{R_3}  \P{1 } $  
 & $+\cL{-}{L_{12}} \cL{-}{L_2} \cL{+}{R_{13}} \cL{+}{R_3}  \P{23} $  
 & $+\cL{+}{L_{13}} \cL{+}{L_{23}} \cL{-}{R_{13}} \cL{-}{R_{23}}  \P{3 } $  
 & $+\cL{+}{L_{13}} \cL{-}{L_{23}} \cL{+}{R_{13}} \cL{-}{R_{23}}  \P{3 } $ \\
   $+\cL{+}{L_{13}} \cL{-}{L_{23}} \cL{-}{R_{13}} \cL{+}{R_{23}}  \P{3 } $  
 & $+\cL{+}{L_{13}} \cL{+}{L_1} \cL{-}{R_{13}} \cL{-}{R_1}  \P{3 } $  
 & $+\cL{+}{L_{13}} \cL{-}{L_1} \cL{-}{R_{13}} \cL{+}{R_1}  \P{3 } $  
 & $+\cL{-}{L_{13}} \cL{+}{L_{23}} \cL{+}{R_{13}} \cL{-}{R_{23}}  \P{3 } $ \\
   $+\cL{-}{L_{13}} \cL{+}{L_{23}} \cL{-}{R_{13}} \cL{+}{R_{23}}  \P{3 } $  
 & $+\cL{-}{L_{13}} \cL{-}{L_{23}} \cL{+}{R_{13}} \cL{+}{R_{23}}  \P{3 } $  
 & $+\cL{-}{L_{13}} \cL{+}{L_1} \cL{+}{R_{13}} \cL{-}{R_1}  \P{3 } $  
 & $+\cL{-}{L_{13}} \cL{-}{L_1} \cL{+}{R_{13}} \cL{+}{R_1}  \P{3 } $ \\
   $+\cL{-}{L_{13}} \cL{+}{L_3} \cL{+}{R_{12}} \cL{-}{R_2}  \P{1 } $  
 & $+\cL{-}{L_{13}} \cL{-}{L_3} \cL{+}{R_{12}} \cL{+}{R_2} \bP{23} $ 
 & $+\cL{+}{L_{23}} \cL{+}{L_2} \cL{-}{R_{23}} \cL{-}{R_2}  \P{3 } $  
 & $+\cL{+}{L_{23}} \cL{-}{L_2} \cL{-}{R_{23}} \cL{+}{R_2}  \P{3 } $ \\
   $+\cL{+}{L_{23}} \cL{+}{L_3} \cL{-}{R_{12}} \cL{-}{R_1}  \P{2 } $  
 & $+\cL{+}{L_{23}} \cL{-}{L_3} \cL{-}{R_{12}} \cL{+}{R_1} \bP{13} $  
 & $+\cL{-}{L_{23}} \cL{+}{L_2} \cL{+}{R_{23}} \cL{-}{R_2}  \P{3 } $  
 & $+\cL{-}{L_{23}} \cL{-}{L_2} \cL{+}{R_{23}} \cL{+}{R_2}  \P{3 } $ \\
\end{tabular}} 
\label{w5ssm2a} 
\eeqn

\no $W_5({\rm mixed})$: 
\beqn 
{\begin{tabular}{llll} 
   $\php \h{1 }  \Q{L_2}  \Q{R_2}  \H{1 } \bH{1 } $  
 & $+ \h{1 }  \Q{L_2}  \Q{R_2}  \H{2 } \bH{2 } $ 
 & $+ \h{1 }  \L{L_2}  \L{R_2}  \H{1 } \bH{1 } $  
 & $+ \h{1 }  \L{L_2}  \L{R_2}  \H{2 } \bH{2 } $ \\
   $+ \h{1 } \cL{+}{L_{13}} \cL{-}{R_{13}}  \H{1 } \bH{1 } $  
 & $+ \h{1 } \cL{+}{L_{13}} \cL{-}{R_{13}}  \H{2 } \bH{2 } $  
 & $+ \h{1 } \cL{-}{L_{13}} \cL{+}{R_{13}}  \H{1 } \bH{1 } $  
 & $+ \h{1 } \cL{-}{L_{13}} \cL{+}{R_{13}}  \H{2 } \bH{2 } $ \\
   $+ \h{1 } \cL{+}{L_2} \cL{-}{R_2}  \H{1 } \bH{1 } $  
 & $+ \h{1 } \cL{+}{L_2} \cL{-}{R_2}  \H{2 } \bH{2 } $  
 & $+ \h{1 } \cL{-}{L_2} \cL{+}{R_2}  \H{1 } \bH{1 } $  
 & $+ \h{1 } \cL{-}{L_2} \cL{+}{R_2}  \H{2 } \bH{2 } $ \\
   $+ \h{2 }  \Q{L_1}  \Q{R_1}  \H{1 } \bH{1 } $  
 & $+ \h{2 }  \Q{L_1}  \Q{R_1}  \H{2 } \bH{2 } $  
 & $+ \h{2 }  \L{L_1}  \L{R_1}  \H{1 } \bH{1 } $  
 & $+ \h{2 }  \L{L_1}  \L{R_1}  \H{2 } \bH{2 } $  \\
   $+ \h{2 } \cL{+}{L_{23}} \cL{-}{R_{23}}  \H{1 } \bH{1 } $  
 & $+ \h{2 } \cL{+}{L_{23}} \cL{-}{R_{23}}  \H{2 } \bH{2 } $  
 & $+ \h{2 } \cL{-}{L_{23}} \cL{+}{R_{23}}  \H{1 } \bH{1 } $  
 & $+ \h{2 } \cL{-}{L_{23}} \cL{+}{R_{23}}  \H{2 } \bH{2 } $ \\
   $+ \h{2 } \cL{+}{L_1} \cL{-}{R_1}  \H{1 } \bH{1 } $  
 & $+ \h{2 } \cL{+}{L_1} \cL{-}{R_1}  \H{2 } \bH{2 } $  
 & $+ \h{2 } \cL{-}{L_1} \cL{+}{R_1}  \H{1 } \bH{1 } $  
 & $+ \h{2 } \cL{-}{L_1} \cL{+}{R_1}  \H{2 } \bH{2 } $ 
\end{tabular}}
\label{w5mix2a}
\eeqn
\end{flushleft}

\no $W_5({\rm singlets})$, $W_5({\rm hidden})$: none


}

\def\s{$\pm$}
\def\t{$\mp$}
{\textwidth=7.5in
\oddsidemargin=-18mm
\topmargin=-5mm
\renewcommand{\baselinestretch}{1.3}


\begin{flushleft}
\begin{table}
{\rm \large\bf C Left--Right~Symmetric~Model~2~Fields}
\begin{eqnarray*}
\begin{tabular}{|c|c|c|rrrrrr|c|rr|}
\hline
  $F$ & SEC & $SU(3)_C\times SU(2)_{L;R;C}$
   & $Q_{1^\prime}$ & $Q_{2^\prime}$ & $Q_{3^\prime}$   
   & $Q_4$ & $Q_5$ & $Q_6$ 
   & $SU(4)_{H_{1;2}}$ 
   & $Q_{7^{\prime\prime}}$ & $Q_{8^{\prime\prime}}$ \\
\hline
   $Q_{L_1}$ & $\mb_1\oplus$ & $(3,2,1,1)$ & 
	  -2 &   -2 &   -4 &   -2 &    0 &    0 &
   $(1,1)$&    0 &  -16 \\
   $Q_{R_1}$ & $\mb_1+\mzeta+2\mgamma$           & $({\bar3},1,2,1)$ & 
           2 &    2 &    4 &    2 &    0 &    0 &
   $(1,1)$&    0 &   16 \\
   $L_{L_1}$ &            & $(1,2,1,2)$ & 
          -2 &   -2 &    4 &    2 &    0 &    0 &
   $(1,1)$ &  -8 &   16 \\
   $L_{R_1}$ &            & $(1,1,2,2)$ & 
	   2 &    2 &   -4 &   -2 &    0 &    0 &
   $(1,1)$ &   8 &  -16 \\
   $S_{L_1}$ &            & $(1,2,1,1)$ & 
	  -2 &   -2 &    4 &    2 &    0 &    0 &
   $(1,1)$ &  16 &   16 \\
   $S_{R_1}$ &            & $(1,1,2,1)$ & 
	   2 &    2 &   -4 &   -2 &    0 &    0 &
   $(1,1)$ & -16 &  -16 \\
\hline
   $Q_{L_2}$ & $\mb_2\oplus$ & $(3,2,1,1)$ & 
	   2 &   -2 &   -4 &    0 &   -2 &    0 &
   $(1,1)$   &   0 &  -16 \\
\hline
\end{tabular}
\label{model2matter1}
\end{eqnarray*}
Table 2: Model 2 fields. Charges have been multiplied by 4.
\end{table}
\end{flushleft}

\begin{flushleft}
\begin{table}
\begin{eqnarray*}
\begin{tabular}{|c|c|c|rrrrrr|c|rr|}
\hline
  $F$ & SEC & $SU(3)_C\times SU(2)_{L;R;C}$
   & $Q_{1^\prime}$ & $Q_{2^\prime}$ & $Q_{3^\prime}$   
   & $Q_4$ & $Q_5$ & $Q_6$ 
   & $SU(4)_{H_{1;2}}$ 
   & $Q_{7^{\prime\prime}}$ & $Q_{8^{\prime\prime}}$ \\
\hline  
  ${ D}_{1}$ & $I+S+2\mgamma$ & $({3},1,1,1)$ & 
	  -4 &   -4 &    0 &    0 &    0 &    0 & 
   $(1,1)$   &   16 &    0  \\
${\bar D}_{1}$&              & $({\bar 3},1,1,1)$ & 
	   4 &    4 &    0 &    0 &    0 &    0 & 
   $(1,1)$   &  -16 &    0  \\
   ${ D}_{2}$&               & $({3},1,1,1)$ & 
	   4 &   -4 &    0 &    0 &    0 &    0 &
   $(1,1)$   &   16 &    0\\
   ${\bar D}_{2}$ &               & $({\bar 3},1,1,1)$ & 
	  -4 &    4 &    0 &    0 &    0 &    0 & 
   $(1,1)$   &  -16 &    0 \\ 
   ${ D}_{3}$&               & $({3},1,1,1)$ & 
	   0 &    8 &    0 &    0 &    0 &    0 & 
   $(1,1)$   &   16 &    0 \\
   ${\bar D}_{3}$ &               & $({\bar 3},1,1,1)$ & 
           0 &   -8 &    0 &    0 &    0 &    0 &
   $(1,1)$   &  -16 &    0 \\ 
\hline 
   $      D_{\malpha\mbeta}$  &$\xi=S+$             & $(   3,1,1,1)$ & 
	   0 &    4 &    8 &    0 &    0 &    0 &
   $(1,1)$   &    0 &   32  \\
\hline
\end{tabular}
\label{model2matter2}
\end{eqnarray*}
Table 2 continued: Model 2 fields.
\end{table}
\end{flushleft}

\hfill\vfill\eject}


\setcounter{section}{3}

{\textwidth=7.5in
\oddsidemargin=-18mm
\topmargin=-5mm
\renewcommand{\baselinestretch}{1.3}

\section{Left--Right Symmetric Model 2 Superpotential Terms} 
\begin{flushleft}
\no $W_3({\rm observable})$:
\beqn 
{\begin{tabular}{llllll}
   $\php\h{1 }  \h{2 }  \P{3 } $  
 & $+\h{1 }  \Q{L_1}  \Q{R_1} $  
 & $+\h{1 }  \L{L_1}  \L{R_1} $  
 & $+\h{1 }  \S{L_1}  \S{R_1} $  
 & $+\h{1 }  \cl{L_1} \bcl{R_1} $  
 & $+\h{1 } \bcl{L_1}  \cl{R_1} $\\
   $+\h{2 }  \Q{L_2}  \Q{R_2} $  
 & $+\h{2 }  \L{L_2}  \L{R_2} $  
 & $+\h{2 }  \S{L_2}  \S{R_2} $  
 & $+\h{2 }  \cl{L_2} \bcl{R_2} $  
 & $+\h{2 } \bcl{L_2}  \cl{R_2} $  
 & $+\Q{L_1} \bD{1 }  \S{L_1} $\\
   $+\Q{L_2} \bD{2 }  \S{L_2} $  
 & $+\Q{L_3} \bD{3 }  \S{L_3} $  
 & $+\Q{R_1}  \D{1 }  \S{R_1} $  
 & $+\Q{R_2}  \D{2 }  \S{R_2} $  
 & $+\Q{R_3}  \D{3 }  \S{R_3} $  
 & $+\D{\ab} \bD{\ab}  \P{3 } $\\
   $+\D{\ab} \bD{3 }  \p{4 } $  
 & $+\D{1 } \bD{2 } \bX{12} $  
 & $+\D{1 } \bD{3 } \bX{31} $  
 & $+\D{2 } \bD{1 }  \X{12} $  
 & $+\D{2 } \bD{3 } \bX{23} $  
 & $+\D{3 } \bD{\ab} \bp{4 } $\\
   $+\D{3 } \bD{1 }  \X{31} $  
 & $+\D{3 } \bD{2 }  \X{23} $  
 & $+\cl{L_1} \bcl{L_1}  \P{1 } $  
 & $+\bcl{L_1} \bcl{L_2} \bp{4 } $  
 & $+\cl{L_2} \bcl{L_2}  \P{2 } $  
 & $+\cl{R_1} \bcl{R_1}  \P{1 } $\\
   $+\cl{R_1}  \cl{R_2}  \p{4 } $  
 & $+\cl{R_2} \bcl{R_2}  \P{2 } $  
 & $+\N{\ab} \bN{\ab}  \P{3 } $  
\end{tabular}} 
\label{w3ssm} 
\eeqn

\no $W_3({\rm singlets})$:
\beqn 
{\begin{tabular}{llllll}
   $\php\P{3 }\ps{2}{+} \bps{2}{+} $  
 & $+\P{3 }  \ps{2}{-}  \bps{2}{-} $  
 & $+\P{3 }  \ps{1}{+}  \bps{1}{+} $  
 & $+\P{3 }  \ps{1}{-}  \bps{1}{-} $  
 & $+\P{3 }  \ps{3}{+}  \bps{3}{+} $  
 & $+\P{3 }  \ps{3}{-}  \bps{3}{-} $\\
   $+\P{3 }  \p{4 }     \bp{4 } $  
 & $+\X{31}  \bX{23}    \bX{12} $  
 & $+\bX{31} \X{23}     \X{12} $  
 & $+\X{12}  \ps{2}{+}  \ps{2}{-} $  
 & $+\X{12}  \ps{1}{+}  \ps{1}{-} $  
 & $+\X{12}  \ps{3}{+}  \ps{3}{-} $\\
   $+\bX{12} \bps{2}{+} \bps{2}{-} $  
 & $+\bX{12} \bps{1}{+} \bps{1}{-} $  
 & $+\bX{12} \bps{3}{+} \bps{3}{-} $  
\end{tabular}} 
\label{w3sig} 
\eeqn

\no $W_3({\rm hidden})$: none
\vskip .8truecm

\no $W_4({\rm observable})$:
\beqn 
{\begin{tabular}{lllll}
   $\php \Q{L_1}  \Q{L_1}  \D{\ab}  \ps{1}{+} $  
 & $+ \Q{L_1}  \Q{L_3}  \Q{R_1}  \Q{R_3} $  
 & $+ \Q{L_1}  \Q{R_1}  \L{L_3}  \L{R_3} $  
 & $+ \Q{L_1}  \Q{R_1}  \S{L_3}  \S{R_3} $  
 & $+ \Q{L_2}  \Q{L_2}  \D{\ab} \bps{2}{-} $ \\
   $+ \Q{L_2}  \Q{L_3}  \Q{R_2}  \Q{R_3} $  
 & $+ \Q{L_2}  \Q{R_2}  \L{L_3}  \L{R_3} $  
 & $+ \Q{L_2}  \Q{R_2}  \S{L_3}  \S{R_3} $  
 & $+ \Q{L_3}  \Q{R_3}  \L{L_1}  \L{R_1} $  
 & $+ \Q{L_3}  \Q{R_3}  \L{L_2}  \L{R_2} $ \\
   $+ \Q{L_3}  \Q{R_3}  \S{L_1}  \S{R_1} $  
 & $+ \Q{L_3}  \Q{R_3}  \S{L_2}  \S{R_2} $  
 & $+ \Q{L_3}  \Q{R_3}  \cl{L_1} \bcl{R_1} $  
 & $+ \Q{L_3}  \Q{R_3} \bcl{L_1}  \cl{R_1} $  
 & $+ \Q{L_3}  \Q{R_3}  \cl{L_2} \bcl{R_2} $ \\
   $+ \Q{L_3}  \Q{R_3} \bcl{L_2}  \cl{R_2} $  
 & $+ \Q{R_1}  \Q{R_1} \bD{\ab} \bps{1}{+} $  
 & $+ \Q{R_2}  \Q{R_2} \bD{\ab}  \ps{2}{-} $  
 & $+ \L{L_1}  \L{L_1}  \ps{1}{-}  \p{4 } $  
 & $+ \L{L_1}  \L{L_3}  \L{R_1}  \L{R_3} $ \\
   $+ \L{L_1}  \L{R_1}  \S{L_3}  \S{R_3} $  
 & $+ \L{L_2}  \L{L_2} \bps{2}{+}  \p{4 } $  
 & $+ \L{L_2}  \L{L_3}  \L{R_2}  \L{R_3} $  
 & $+ \L{L_2}  \L{R_2}  \S{L_3}  \S{R_3} $  
 & $+ \L{L_3}  \L{R_3}  \S{L_1}  \S{R_1} $ \\
   $+ \L{L_3}  \L{R_3}  \S{L_2}  \S{R_2} $  
 & $+ \L{L_3}  \L{R_3}  \cl{L_1} \bcl{R_1} $  
 & $+ \L{L_3}  \L{R_3} \bcl{L_1}  \cl{R_1} $  
 & $+ \L{L_3}  \L{R_3}  \cl{L_2} \bcl{R_2} $  
 & $+ \L{L_3}  \L{R_3} \bcl{L_2}  \cl{R_2} $ \\
   $+ \L{R_1}  \L{R_1} \bps{1}{-} \bp{4 } $  
 & $+ \L{R_2}  \L{R_2}  \ps{2}{+} \bp{4 } $  
 & $+ \S{L_1}  \S{L_3}  \S{R_1}  \S{R_3} $  
 & $+ \S{L_2}  \S{L_3}  \S{R_2}  \S{R_3} $  
 & $+ \S{L_3}  \cl{L_1}  \S{R_3} \bcl{R_1} $ \\
   $+ \S{L_3} \bcl{L_1}  \S{R_3}  \cl{R_1} $  
 & $+ \S{L_3}  \cl{L_2}  \S{R_3} \bcl{R_2} $  
 & $+ \S{L_3} \bcl{L_2}  \S{R_3}  \cl{R_2} $  
\end{tabular}} 
\label{w4ssm} 
\eeqn

\no $W_4({\rm singlets})$, $W_4({\rm mixed})$, $W_4({\rm hidden})$: none
\newpage

\no $W_5({\rm observable})$:
\beqn 
{\begin{tabular}{llll} 
   $\php \h{1 }  \Q{L_2}  \Q{R_2}  \D{\ab} \bD{\ab} $  
 & $+ \h{1 }  \Q{L_2}  \Q{R_2}  \N{\ab} \bN{\ab} $  
 & $+ \h{1 }  \Q{L_2}  \Q{R_2}  \ps{2}{+} \bps{2}{+} $ 
 & $+ \h{1 }  \Q{L_2}  \Q{R_2}  \ps{2}{-} \bps{2}{-} $ \\ 
   $+ \h{1 }  \Q{L_2}  \Q{R_2}  \ps{1}{+} \bps{1}{+} $  
 & $+ \h{1 }  \Q{L_2}  \Q{R_2}  \ps{1}{-} \bps{1}{-} $  
 & $+ \h{1 }  \Q{L_2}  \Q{R_2}  \ps{3}{+} \bps{3}{+} $ 
 & $+ \h{1 }  \Q{L_2}  \Q{R_2}  \ps{3}{-} \bps{3}{-} $ \\ 
   $+ \h{1 }  \Q{L_2}  \Q{R_2}  \p{4 } \bp{4 } $  
 & $+ \h{1 }  \L{L_2}  \L{R_2}  \D{\ab} \bD{\ab} $ 
 & $+ \h{1 }  \L{L_2}  \L{R_2}  \N{\ab} \bN{\ab} $  
 & $+ \h{1 }  \L{L_2}  \L{R_2}  \ps{2}{+} \bps{2}{+} $ \\  
   $+ \h{1 }  \L{L_2}  \L{R_2}  \ps{2}{-} \bps{2}{-} $  
 & $+ \h{1 }  \L{L_2}  \L{R_2}  \ps{1}{+} \bps{1}{+} $ 
 & $+ \h{1 }  \L{L_2}  \L{R_2}  \ps{1}{-} \bps{1}{-} $  
 & $+ \h{1 }  \L{L_2}  \L{R_2}  \ps{3}{+} \bps{3}{+} $ \\ 
   $+ \h{1 }  \L{L_2}  \L{R_2}  \ps{3}{-} \bps{3}{-} $  
 & $+ \h{1 }  \L{L_2}  \L{R_2}  \p{4 } \bp{4 } $ 
 & $+ \h{1 }  \D{\ab} \bD{\ab}  \S{L_2}  \S{R_2} $  
 & $+ \h{1 }  \D{\ab} \bD{\ab}  \cl{L_2} \bcl{R_2} $ \\  
   $+ \h{1 }  \D{\ab} \bD{\ab} \bcl{L_2}  \cl{R_2} $  
 & $+ \h{1 }  \S{L_2}  \S{R_2}  \N{\ab} \bN{\ab} $ 
 & $+ \h{1 }  \S{L_2}  \S{R_2}  \ps{2}{+} \bps{2}{+} $  
 & $+ \h{1 }  \S{L_2}  \S{R_2}  \ps{2}{-} \bps{2}{-} $ \\  
   $+ \h{1 }  \S{L_2}  \S{R_2}  \ps{1}{+} \bps{1}{+} $  
 & $+ \h{1 }  \S{L_2}  \S{R_2}  \ps{1}{-} \bps{1}{-} $ 
 & $+ \h{1 }  \S{L_2}  \S{R_2}  \ps{3}{+} \bps{3}{+} $  
 & $+ \h{1 }  \S{L_2}  \S{R_2}  \ps{3}{-} \bps{3}{-} $ \\ 
   $+ \h{1 }  \S{L_2}  \S{R_2}  \p{4 } \bp{4 } $  
 & $+ \h{1 }  \cl{L_2} \bcl{R_2}  \N{\ab} \bN{\ab} $ 
 & $+ \h{1 }  \cl{L_2} \bcl{R_2}  \ps{2}{+} \bps{2}{+} $  
 & $+ \h{1 }  \cl{L_2} \bcl{R_2}  \ps{2}{-} \bps{2}{-} $ \\
\end{tabular}} 
\nolabel
\eeqn
\newpage

\no $W_5({\rm observable})$ continued:
\beqn 
{\begin{tabular}{llll} 
   $+ \cl{L_1} \bcl{L_2} \bcl{R_1}  \cl{R_2}  \P{3 } $  
 & $+\bcl{L_1}  \cl{L_2} \bcl{L_2} \bcl{L_2} \bp{4 } $  
 & $+\bcl{L_1}  \cl{L_2}  \cl{R_1} \bcl{R_2}  \P{3 } $  
 & $+\bcl{L_1}  \cl{L_2} \bcl{R_2} \bcl{R_2} \bp{4 } $ \\
   $+\bcl{L_1} \bcl{L_2}  \cl{R_1} \bcl{R_1} \bp{4 } $  
 & $+\bcl{L_1} \bcl{L_2}  \cl{R_1}  \cl{R_2}  \P{3 } $  
 & $+\bcl{L_1} \bcl{L_2}  \cl{R_2} \bcl{R_2} \bp{4 } $  
 & $+\bcl{L_1} \bcl{L_2}  \N{\ab} \bN{\ab} \bp{4 }   $ \\
   $+\bcl{L_1} \bcl{L_2}  \ps{2}{+} \bps{2}{+} \bp{4 } $  
 & $+\bcl{L_1} \bcl{L_2}  \ps{2}{-} \bps{2}{-} \bp{4 } $  
 & $+\bcl{L_1} \bcl{L_2}  \ps{1}{+} \bps{1}{+} \bp{4 } $  
 & $+\bcl{L_1} \bcl{L_2}  \ps{1}{-} \bps{1}{-} \bp{4 } $ \\
   $+\bcl{L_1} \bcl{L_2}  \ps{3}{+} \bps{3}{+} \bp{4 } $  
 & $+\bcl{L_1} \bcl{L_2}  \ps{3}{-} \bps{3}{-} \bp{4 } $  
 & $+\bcl{L_1} \bcl{L_2}  \p{4 } \bp{4 } \bp{4 } $  
 & $+ \cl{L_2}  \cl{L_2}  \cl{R_1} \bcl{R_2}  \p{4 } $ \\
   $+ \cl{L_2} \bcl{L_2}  \cl{R_1}  \cl{R_2}  \p{4 } $  
 & $+ \cl{R_1}  \cl{R_1} \bcl{R_1}  \cl{R_2}  \p{4 } $  
 & $+ \cl{R_1}  \cl{R_2}  \cl{R_2} \bcl{R_2}  \p{4 } $  
 & $+ \cl{R_1}  \cl{R_2}  \N{\ab} \bN{\ab}  \p{4 } $ \\ 
   $+ \cl{R_1}  \cl{R_2}  \ps{2}{+} \bps{2}{+}  \p{4 } $  
 & $+ \cl{R_1}  \cl{R_2}  \ps{2}{-} \bps{2}{-}  \p{4 } $ 
 & $+ \cl{R_1}  \cl{R_2}  \ps{1}{+} \bps{1}{+}  \p{4 } $  
 & $+ \cl{R_1}  \cl{R_2}  \ps{1}{-} \bps{1}{-}  \p{4 } $ \\ 
   $+ \cl{R_1}  \cl{R_2}  \ps{3}{+} \bps{3}{+}  \p{4 } $  
 & $+ \cl{R_1}  \cl{R_2}  \ps{3}{-} \bps{3}{-}  \p{4 } $ 
 & $+ \cl{R_1}  \cl{R_2}  \p{4 }  \p{4 } \bp{4 } $  
\end{tabular}} 
\label{w5ssm} 
\eeqn

\no $W_5({\rm singlets})$, $W_5({\rm mixed})$, $W_5({\rm hidden})$: none
\end{flushleft}

}

{\textwidth=7.5in
\oddsidemargin=-18mm
\topmargin=-5mm
\renewcommand{\baselinestretch}{1.3}
\smallskip

\begin{flushleft}
\begin{table}
{\rm \large\bf E Left--Right~Symmetric~Model~3~Fields}
\begin{eqnarray*}
\begin{tabular}{|c|c|c|rrrrrrr|c|rrrr|}
\hline

  $F$ & SEC & $(C;L;R)$ & $Q_C$
   & $Q_1$ & $Q_2$ & $Q_3$
   & $Q_4$ & $Q_5$ & $Q_6$
   & $SU(3)_{H_{1;2}}$
   & $Q_7$ & $Q_8$ & $Q_9$ & $Q_{10}$ \\

\hline

  $Q_{L_1}$ & $\mb_1$ & $(3,2,1)$
     &  2 &  2 &  0 &  0 & -2 &  0 &  0 & $(1,1)$        &  0 &  0 &  0 &  0 \\
  $Q_{R_1}$ & & $({\bar3},1,2)$
     & -2 & -2 &  0 &  0 & -2 &  0 &  0 & $(1,1)$        &  0 &  0 &  0 &  0 \\
  $L_{L_1}$ & & $(1,2,1)$
     & -6 &  2 &  0 &  0 & -2 &  0 &  0 & $(1,1)$        &  0 &  0 &  0 &  0 \\
  $L_{R_1}$ & & $(1,1,2)$
     &  6 & -2 &  0 &  0 & -2 &  0 &  0 & $(1,1)$        &  0 &  0 &  0 &  0 \\

\hline

  $Q_{L_2}$ & $\mb_2$ & $(3,2,1)$
     &  2 &  0 &  2 &  0 &  0 & -2 &  0 & $(1,1)$        &  0 &  0 &  0 &  0 \\
  $Q_{R_2}$ & & $({\bar3},1,2)$
     & -2 &  0 & -2 &  0 &  0 & -2 &  0 & $(1,1)$        &  0 &  0 &  0 &  0 \\
  $L_{L_2}$ & & $(1,2,1)$
     & -6 &  0 &  2 &  0 &  0 & -2 &  0 & $(1,1)$        &  0 &  0 &  0 &  0 \\
\hline
\end{tabular}
\nolabel
\end{eqnarray*}
Table 3: Model 3 fields.
\end{table}
\end{flushleft}

\begin{flushleft}
\begin{table}
\begin{eqnarray*}
\begin{tabular}{|c|c|c|rrrrrrr|c|rrrr|}
\hline

  $F$ & SEC & $(C;L;R)$ & $Q_C$
   & $Q_1$ & $Q_2$ & $Q_3$
   & $Q_4$ & $Q_5$ & $Q_6$
   & $SU(3)_{H_{1;2}}$
   & $Q_7$ & $Q_8$ & $Q_9$ & $Q_{10}$ \\

\hline

$\p{1 }$& $\mS+\mb_1+\mb_2$  & $(1,1,1)$
     &  0 &  0 &  0 & -2 & -2 & -2 &  0 & $(1,1)$        & -2 &  0 &  6 &  0 \\
$\p{2 }$& $+\malpha+2\mgamma$ & $(1,1,1)$
     &  0 &  0 &  0 &  2 & -2 & -2 &  0 & $(1,1)$        &  2 &  0 & -6 &  0 \\

\hline

$\p{3 }$ & $\mS+\mb_1+\mb_3$  & $(1,1,1)$
     &  0 &  0 & -2 &  0 & -2 &  0 & -2 & $(1,1)$        & -2 &  0 &  6 &  0 \\
$\p{4 }$ & $+\malpha+2\mgamma$ & $(1,1,1)$
     &  0 &  0 &  2 &  0 & -2 &  0 & -2 & $(1,1)$        &  2 &  0 & -6 &  0 \\

\hline
$\p{5 }$ & $\mS+\mb_2+\mb_3$  & $(1,1,1)$
     &  0 & -2 &  0 &  0 &  0 & -2 & -2 & $(1,1)$        & -2 &  0 &  6 &  0 \\
$\p{6 }$ & $+\malpha+2\mgamma$ & $(1,1,1)$
     &  0 &  2 &  0 &  0 &  0 & -2 & -2 & $(1,1)$        &  2 &  0 & -6 &  0 \\

\hline

$\p{7 }$ & $\mS+\mb_1+\mb_2$ & $(1,1,1)$
     &  0 &  2 & -2 &  0 &  0 &  0 &  0 & $(1,1)$        &  4 &  0 &  0 &  0 \\
\hline
\end{tabular}
\nolabel
\end{eqnarray*}
Table 3 continued: Model 3 fields.
\end{table}
\end{flushleft}


\begin{flushleft}
\begin{table}
\begin{eqnarray*}
\begin{tabular}{|c|c|c|rrrrrrr|c|rrrr|}
\hline

  $F$ & SEC & $(C;L;R)$ & $Q_C$
   & $Q_1$ & $Q_2$ & $Q_3$
   & $Q_4$ & $Q_5$ & $Q_6$
   & $SU(3)_{H_{1;2}}$
   & $Q_7$ & $Q_8$ & $Q_9$ & $Q_{10}$ \\

\hline
$\xH{+}{1}$ & $\bo+\mS+\mb_1$ & $(1,1,1)$
     &  0 &  2 &  0 &  0 &  0 & -2 & -2 & $(3,1)$        &  0 &  2 &  0 & -2 \\
$\xH{-}{1}$ & $+\malpha+2\mgamma$ & $(1,1,1)$
     &  0 & -2 &  0 &  0 &  0 & -2 & -2 & $({\bar3},1)$  &  0 & -2 &  0 &  2 \\

\hline

$\xH{+}{2}$ & $\bo+\mS+\mb_2$ & $(1,1,1)$
     &  0 &  0 &  2 &  0 & -2 &  0 & -2 & $(3,1)$        &  0 &  2 &  0 & -2 \\
$\xH{-}{2}$ & $+\malpha+2\mgamma$ & $(1,1,1)$
     &  0 &  0 & -2 &  0 & -2 &  0 & -2 & $({\bar3},1)$  &  0 & -2 &  0 &  2 \\

\hline

$\xH{+}{3}$ & $\bo+\mS+\mb_3$ & $(1,1,1)$
     &  0 &  0 &  0 &  2 & -2 & -2 &  0 & $(3,1)$        &  0 &  2 &  0 & -2 \\
$\xH{-}{3}$ & $+\malpha+2\mgamma$ & $(1,1,1)$
     &  0 &  0 &  0 & -2 & -2 & -2 &  0 & $({\bar3},1)$  &  0 & -2 &  0 &  2 \\

\hline
$\H{4}$ & $\bo+\mS+\mb_1$ & $(1,1,1)$
     & -3 &  1 & -1 & -1 &  0 &  0 &  0 & $(3,1)$        &  0 & -1 & -3 & -2 \\
\hline
\end{tabular}
\label{chiralmatter}
\end{eqnarray*}
Table 3 continued: Model 3 fields.
\end{table}
\end{flushleft}

\hfill\vfill\eject}


}

\newpage


\vfill\eject

{{\oddsidemargin  10.5pt \evensidemargin  10.5pt
\textheight  612pt \textwidth  432pt
\headheight  12pt \headsep  20pt
\footheight  12pt \footskip  40pt

\bibliographystyle{unsrt}
}}
\vfill\eject

\end{document}